\documentclass[pra,preprint,showpacs]{revtex4}
\usepackage[centertags]{amsmath}
\usepackage{amsfonts}
\usepackage{amssymb}
\usepackage{amsthm} 
\usepackage{newlfont}
\usepackage{epsfig}
\usepackage{amscd}
\usepackage{graphicx}
\usepackage{epsfig}
\usepackage{amscd}
\usepackage{color}

\newcommand{\beq}{\begin{equation}}
\newcommand{\eeq}{\end{equation}}
\newcommand{\ba}{\begin{array}}
\newcommand{\ea}{\end{array}}
\newcommand{\bea}{\begin{eqnarray}}
\newcommand{\eea}{\end{eqnarray}}
\begin{document}

\begin{center}
{\large \sc \bf { 
Temperature-dependent remote control of polarization and 
 coherence intensity  with  sender's pure initial state
}}

\vskip 15pt

{\large 
E.B. Fel'dman, E.I. Kuznetsova, and A.I. Zenchuk  
}

\vskip 8pt

{\it Institute of Problems of Chemical Physics, RAS,
Chernogolovka, Moscow reg., 142432, Russia},\\
% e-mail:  efeldman@icp.ac.ru, zenchuk@itp.ac.ru 

\end{center}

%\today

\begin{abstract}
We study the remote creation of  the polarization and intensity of the first-order 
coherence (or coherence intensity) in long spin-1/2 chains with 
one qubit sender and receiver. Therewith we use a physically motivated  initial 
condition with the pure state of the sender and the 
thermodynamical equilibrium state of the other  nodes. 
The main part of the creatable region is a one-to-one map of the initial-state (control)
parameters, except  the small subregion twice covered by the control parameters, which appears owing to the 
chosen initial state. 
The polarization and coherence intensity  behave differently in the  state creation process.
In particular, the coherence intensity  can not  reach  any significant value  unless the   polarization 
is large in long chains 
(unlike the short ones), but the opposite is not true.
The  coherence intensity 
vanishes with an increase in the
chain length, while the polarization (by absolute value) is not sensitive to this parameter.
We represent several characteristics of the creatable polarization  and coherence intensity
and describe their relation to 
the parameters of the initial state.
The link to the eigenvalue-eigenvector  parametrization  of the receiver's state-space 
 is given.

 \end{abstract}

\maketitle

%%%%%%%%%%%%%%%%
\section{Introduction}
\label{Sec:Introduction}

The problem of remote state creation is one of the problems attracting attention of many scientists. 
It goes back to 
the problem  of arbitrary state transfer \cite{Bose}.  
Being simply formulated, this problem faces many obstacles for its practical realization, such as the noise, 
the interaction with environment 
(in particular, with the remote nodes which are not taken into account in many proposed algorithms).
Because of these obstacles, the characteristics  obtained in the 
original  algorithms of   state transfer along  the either   fully engineered  \cite{CDEL,ACDE,KS}  
or   boundary-controlled  \cite{GKMT,WLKGGB} spin-1/2 chain 
are hard reachable in practical realizations \cite{CRMF,ZASO,ZASO2,ZASO3,SAOZ}. In particular, 
the state-transfer  probability becomes significantly  less 
then one. Therefore, 
it is reasonable to replace 
the perfect state transfer by the high probability state transfer { \cite{KZ_2008}}  
({in particular}, using 
the optimized boundary-controlled chain 
 \cite{NJ,SAOZ}), which is much simpler for realization. 
 Using this concept, a set of effects has been studied in both homogeneous and 
engineered spin chains, for instance, 
 the thermal effect on a state transfer \cite{BK}, 
the mixed state transfer \cite{C}. Moreover, being simply
realizable, the homogeneous chains  acquire significance in the algorithms of the multidimensional
state transfer   \cite{QWL,B}, long-time  high probability state transfer  
\cite{JKSS,SO}, multiple-channel state transfer 
\cite{BB,SJBB}.  Among other models, we mention the quantum state transfer in an engineered  chain of 
superconducting quantum bits \cite{SKKVP}.
 
Nevertheless, the algorithms of perfect state transfer (allowing us 
to restore exactly  the initial state of sender at the receiver site) 
are very 
attractive due to the simple and illustrative ideas they are based on. 
The first idea is the realization  of the transfered state as a superposition of 
functions oscillating in time with rational (integer) frequencies \cite{CDEL,ACDE,KS}, 
so that the transfered state reproduces exactly the initial sender's state at some time instant. 
The second one is the  selective choice of the active modes
(i.e., only a few of the above mentioned oscillating functions
have large amplitudes, therewith  their frequencies are not rational numbers)
realizing the state transfer 
\cite{GKMT,WLKGGB,NJ,SAOZ}. 
Both these effects are used (either explicitly or implicitly) in many contemporary 
state transfer algorithms \cite{LS,BBVB,QWZ}. 

However, working with the high-probability state transfer stimulates 
the intention to get advantages  from  the apparently destructive 
effects of  spin dynamics. A way to realize this intention is based on the observation that  the 
non-unit probability of a pure state transfer means that we deal with the mixed state of the receiver rather then
with the pure one.
This fact prompts us  to replace  the pure state transfer
with the   mixed state creation thus establishing  a map between 
the parameters of the initial sender state and those of the receiver state,
which was first realized in photon systems 
\cite{ZZHE,BPMEWZ,BBMHP,PBGWK2,PBGWK,DLMRKBPVZBW,XLYG,PSB}.
Therewith,  all the parameters of the mixed receiver's state are  the
parameters of potential interest  in realization of quantum algorithms. {
In some sense,  the remote state creation can be 
referred to as the information transfer along a chain \cite{YBB,Z_2012,PS}.}

In the recent papers, a new algorithm of the remote state creation 
in a spin-1/2 chain with an either  mixed  
tensor product  \cite{Z_2014} or  pure \cite{BZ_2015} initial state
is considered, where 
 the creatable region of the whole state-space of the one-qubit receiver is explicitly shown. 
In that state-creation algorithm,  the parameters of sender's 
initial state cover the whole sender's state space 
and can take any value from their domain. These parameters are referred to as the control parameters (which 
sometimes 
are supplemented by the time instance) and can be associated with the parameters of a local unitary transformation of the sender. 
Similarly, the  parameters of the receiver's 
state parametrize the 
whole state space. However, in general, 
the creatable  values of these parameters are not arbitrary but they are defined by the both control 
parameters and  evolution of a quantum system. {
As a result, the whole receiver state-space  can not be covered in general.}
Thus, 
the state creation is considered as the following  map:
\begin{eqnarray}\label{cpcp}
{\mbox{control parameters $\to$ creatable parameters}}.
\end{eqnarray}
The choice of a preferable parametrization of the both sender's initial state and receiver's creatable state 
is an important step in characterization of  the state creation algorithm. In the above references, 
the parameters of a local unitary transformation of the sender are considered as control ones, while the list of creatable parameters 
is composed by the independent  eigenvalues and eigenvector-parameters.  
But the disadvantage of those creatable parameters is that they
are not directly measurable.  This prompts us to turn to a physically
detectable parametrization. In our case of the one-qubit receiver 
this parametrization is rather evident. 
First of all we notice that the phase of the non-diagonal element 
is simply creatable by choosing the proper control phase-parameter. 
Thus, disregarding this phase, the receiver's state becomes a  two-parametric one, so that the 
polarization and the intensity of the 
first-order coherence (or coherence intensity for the sake of brevity) 
\cite{FBE,Abragam,BMGP,OBSFA,F}  can be  taken as
suitable quantities parameterizing  the receiver state-space.

In addition, 
along with the physically detectable  parametrization of the creatable region, 
we use the physically motivated initial state of a chain,
which is a pure state of the one-qubit sender along with  the thermodynamic-equilibrium state  of all the other spins.
Thus, the set of  control-parameters  consists of the two types of parameters. The first type involves  two 
independent  parameters of a pure  sender's one-qubit state
(the overall phase is disregarded in advance), which { can be represented in terms of the 
parameters of} a local unitary transformation of the sender. 
The second type is the inverse temperature characterizing the above thermodynamic-equilibrium state, 
this parameter can not be embedded  into the parameters of the  sender's local unitary transformation.  
Although it is not a local parameter of the sender, this macro-parameter  can be used as a 
control parameter governing the state of a system of relatively small size,
which  usually holds  in working with spin systems (thus, there is no need to transfer 
the value of this parameter to the receiver 
side using any additional (classical) communication channel).

As usual, by the state of the receiver we mean the density matrix reduced over all the nodes except for the 
receiver's node, which is  the last node in our model. 
By definition,  the polarization is responsible for the 
diagonal part of the created state (the classical part) while the coherence intensity is associated with the 
non-diagonal part of the state (the quantum part). 
It is demonstrated that these quantities may vanish independently in certain cases.
For instance, 
there is a line of states  in the creatable region with non-zero  polarization and  zero coherence intensity.
There is also another  line  with zero polarization and non-zero coherence intensity. 
However, in long chains,  the significant value of  coherence intensity can be created only together with the 
large value of polarization.  We emphasize that the opposite is not true and  the large 
polarization can appear together with the zero 
coherence intensity in long chains.   We also study the dependence of the discussed properties 
on the control  parameters, in particular, on the temperature responsible for the thermodynamic equilibrium state of a
spin system.

Among the features of our model we point out the following one.  Owing to the chosen initial condition, 
there is a creatable sub-region twice covered by  map 
(\ref{cpcp}). This means  that any state from this subregion can be created 
using the two different  pairs of the  control parameters. 
Although a similar  phenomenon was observed in \cite{Z_2014}, it was not closely 
considered therein. In this paper  we analyze this twice-covered 
subregion (in particular, we describe its boundary) though
it  is rather small in our model.

The paper is organized as follows.
In Sec.\ref{Section:analyt} we give the general description of the considered model involving the
interaction Hamiltonian, the initial state and the associated list of  control parameters.
The physically motivated parametrization of the creatable region in terms of the  polarization and  
coherence intensity 
 is studied in Sec.\ref{Section:coh}. We describe the boundary of the 
creatable region, the one-to-one and  two-to-one mapped creatable subregions  and the 
features of mutual relations between the creatable parameters. 
In Sec.\ref{Section:param1}, we consider the parametrization in terms of the
 eigenvalue and eigenvectors of the creatable state, thus  establishing the link to the 
models considered before.
Basic results are collected in 
Sec.\ref{Section:conclusions}.

%%%%%
\section{Model for numerical simulations of remote state creation}
\label{Section:analyt}
For the purpose of a remote state creation, 
we chose the homogeneous chain of $N$ spin-1/2 particles  
with a pure initial state of the sender (the first node) and the thermal-equilibrium state of the rest system, i.e.,
\begin{eqnarray}\label{initial}
\rho_0 = \left(2\cosh\frac{{b}}{2}\right)^{1-N}(a_0 |0\rangle + a_1 |1\rangle)
(a_0 \langle 0| + a_1^* \langle 1|) \otimes e^{{b} I_{z2}}
\otimes  \dots \otimes e^{{b} I_{zN} },
\end{eqnarray}
where the parameter $b$ is an inverse temperature (more exactly, $b=\frac{\hbar w}{kT}$, where 
$\hbar$ is the Planck constant, $w$ is the Larmor frequency, $k$ is  the 
Boltzmann constant and $T$ is the 
temperature),  $I_{j\alpha}$ ($j=1,\dots,N$, $\alpha=x,y,z$) is the projection of the 
$j$th spin momentum  on the $\alpha$-axis, and
\begin{eqnarray}
a_0=\sin \frac{\alpha \pi}{2},\;\;a_1=e^{2 i \pi \phi} \cos \frac{\alpha \pi}{2},\;\;0\le \phi,\alpha \le 1.
\end{eqnarray}
{ Initial state (\ref{initial}) has  two features  to be clarified.
The first one is  a pure initial state of the first qubit. One could assume that
the mixed initial state (being a more 
general one) has to 
lead to a bigger variety of the receiver's states. However, it was numerically  demonstrated in 
\cite{Z_2014} that  the maximal creatable region corresponds to the pure sender's state, because this is a 
state with 
the  maximal possible eigenvalue (which equals one). 
That observation prompts  us to use a pure initial state in our case as well.
Nevertheless,  we keep in mined that a mixed initial state might be 
preferable for some particular problems especially if we intend to
handle the position and area of the creatable region.
The second point is the initial state of the rest chain which is a uniformly polarized state 
corresponding to the equilibrium  initial state of the spin chain in the 
strong external magnetic field before the spin-spin interaction is turned on.
Alternatively, the equilibrium initial state of 
spins interacting by means of the Hamiltonian (\ref{XY})
could be of interest, but this should not yield the qualitatively different results.  }

Let the dynamics of this chain be governed by the nearest-neighbor XY-Hamiltonian 
\begin{eqnarray}\label{XY}
H=  \sum_{i=1}^{N-1}D (I_{ix} I_{(i+1)x} + I_{iy} I_{(i+1)y}),
\end{eqnarray}
where  $D$ is  the coupling constant between the 
nearest neighbors. 
{ We do not include the interaction with the external 
magnetic field in this Hamiltonian, because the term responsible for this interaction  can be eliminated
passing to the rotating reference frame \cite{Goldman}, which  is possible because our 
Hamiltonian commutes with the $z$-projection operator of the total spin momentum 
(the external magnetic field is $z$-directed).}

Below
we consider the dimensionless time  $\tau=D t$ and write the  evolution of the density  matrix  as
$\rho(\tau)=e^{-i \frac{H}{D} \tau} \rho_0e^{i \frac{H}{D} \tau}$. 
Using the Jourdan-Wigner transformation \cite{JW,CG}, we calculate the state of the last qubit, 
i.e., the reduced density 
matrix
$\rho^N(\tau)=Tr_{1,\dots,N-1} \rho(\tau)$, which reads (see Appendix A for details)

\begin{eqnarray}\label{rhoN0}
\rho_N(\tau)= 
\left(\begin{array}{cc}\displaystyle
\rho^N_{11} & r^N_{12} e^{- 2 i\tilde \Phi_N(\tau) \pi}
 \cr
 r^N_{12} e^{2 i\tilde \Phi_N(\tau) \pi}&
1-\rho^N_{11}
\end{array}
\right),
\end{eqnarray}
where 
\begin{eqnarray}\label{rho11}
&&
\rho^N_{11}= \frac{1}{2}\left(1-R_N^2 \cos(\alpha\pi)+(1-R_N^2) \tanh \frac{{b}}{2}\right),
\\\label{r12}
&&r^N_{12}=\frac{1}{2}R_N \sin (\alpha \pi) \left(\tanh\frac{{b}}{2}\right)^{N-1},\;\; \\\label{tPhi}
&&\tilde \Phi_N(\tau) =   \Phi_N(\tau)+\phi + 
 \frac{1}{2} (N-1).
\end{eqnarray}
Here the  functions  $R_N$ and $\Phi_N$ are the 
scalar evolution characteristics of the whole  transmission line.
The derivation of formula (\ref{rhoN0}) 
is
given in  Appendix A, eqs.(\ref{f},\ref{fRPhi}), 
where 
the functions  $R_N$ and $\Phi_N$ are obtained as 
the amplitude and the phase of some
function $f_N(\tau)$  coinciding  with the transition amplitude for the model 
of a pure state transfer  along a homogeneous chain  \cite{CDEL}. 
Emphasize that the both $R_N$ and $\Phi_N$ 
do not depend on the parameters of the system's 
initial state $\alpha$, ${b}$ and 
$\phi$. These parameters appear explicitly in formulas (\ref{rhoN0}-\ref{tPhi}) and are referred to as the
control parameters. They  can take arbitrary  values, unlike the 
 creatable parameters, which characterize the receiver's
 state   and are defined by   both the control parameters and evolution of the system. 

Apparently, eq.(\ref{tPhi}) means that any phase $\tilde \Phi_N$ at the required instant $\tau$
can be created taking a proper value of the control parameter 
$\phi$: $\phi=\tilde \Phi_N(\tau) -\Phi_N(\tau)-
 \frac{1}{2}(N-1)$. Thus, the $\tilde \Phi_N$-creation becomes a trivial task
 and is disregarded below. In other words, we consider 
the creation of $\rho^N_{11}$ and $r^N_{12}$ in the density matrix (\ref{rhoN0})
using the two control parameters $\alpha$ and  ${b}$. Therefore,   
matrix (\ref{rhoN0}) should be replaced  with the following one:
\begin{eqnarray}\label{rhoN}
\hat \rho^N(\tau)= 
\left(\begin{array}{cc}\displaystyle
\rho^N_{11} & r^N_{12} 
 \cr
 r^N_{12}&
1-\rho^N_{11}
\end{array}
\right).
\end{eqnarray}

Instead of describing the receiver's creatable region directly  in terms of the elements of 
 density matrix (\ref{rhoN}), we  
 use a 
proper parametrization of the density matrix, i.e., we introduce 
a pair of so-called creatable parameters whose domain covers the whole receiver's state 
space. The requirement to this parametrization is that it must clearly separate 
the part of the whole state space, 
which can be created varying the control parameters  
(the creatable region) from 
the part, which can not be created in this way (the unavailable region).
In addition, the preference should be given to the parametrization by  the physically detectable parameters.
Accordingly, in Sec.\ref{Section:coh} we concentrate on the parametrization in terms of the
polarization and coherence intensity \cite{Abragam,BMGP,OBSFA,F}. Both of these parameters  can be 
measured in an experiment  
(physical parametrization) and allow us to  separate the creatable region from the unavailable one.

\section{ 
Polarization and coherence intensity as measurable parameters of creatable state }
\label{Section:coh}
Hereafter   we consider the
physically motivated parametrization of the one-qubit receiver's state in terms of 
the polarization $I$ and the coherence intensity $J$. These 
parameters  are related with the  elements of the density matrix 
$\hat \rho^N$ (\ref{rhoN}) as follows:
\begin{eqnarray}\label{IJ0}
&&
I(\tau)= {{\mbox Tr}}(\hat\rho^N(\tau) I_{zN}) =  \rho_{11}(\tau) -\frac{1}{2} =
\frac{1}{2}\left((1-R_N^2(\tau))\tanh\frac{{b}}{2}-R_N^2(\tau) \cos (\alpha \pi) \right)
,\\\label{Jdef0}
&&
J(\tau) =  | \rho_{12}(\tau)|^2= 
\frac{1}{4} R_N^2(\tau) \sin^2(\alpha\pi)  \left(\tanh\frac{{b}}{2}\right)^{2(N-1)}.
\end{eqnarray}
In terms of $I$ and $J$, the density matrix $\hat\rho^N$ (\ref{rhoN}) reads
\begin{eqnarray}\label{rhoN3}
\hat\rho^N(\tau)=\frac{1}{2} E + \left(
\begin{array}{cc}
 I(\tau)&\sqrt{J(\tau)}\cr
\sqrt{J(\tau)} & -I(\tau)
\end{array}
\right),
\end{eqnarray}
where $E$ is the $2\times 2$ identity matrix.
The positivity of the density matrix $\hat\rho^N$ gives  the only constraint on the admissable
values of $I$ and $J$, 
\begin{eqnarray}\label{boundary}
I^2 + J \le \frac{1}{4}.
\end{eqnarray}
Inequality (\ref{boundary}) specifies the whole receiver's state-space.
The parameters $I$ and $J$ in formula (\ref{rhoN3}) characterize the two quite different 
types of deviation of the density matrix
from the 
state $\frac{1}{2} E$, which are called  the classical (the polarization $I$) and the 
quantum (the coherence intensity $J$) deviations. Both depend not only on the control 
parameters $\alpha$ and ${b}$, but also 
on the absolute value  $R_N$ of the transition amplitude
governing the time evolution  of the creatable region.
We assume the positivity of $b$, which is physically justified. The case of negative  $b$ is equivalent to the
replacement $(I,\alpha) \to (-I,\alpha \pm  1)$ and is not considered here.

Before proceeding to
 study of the creatable region,  it is worthwhile  to
describe the amplitude  $R_N$  as a function of the chain length $N$ in more details.

%%%%%%%%%%%%%%%%%%%
\subsection{Amplitude  $R_N$ as a global evolution characteristics of transmission line}
\label{Section:RN}
\begin{figure*}
\epsfig{file=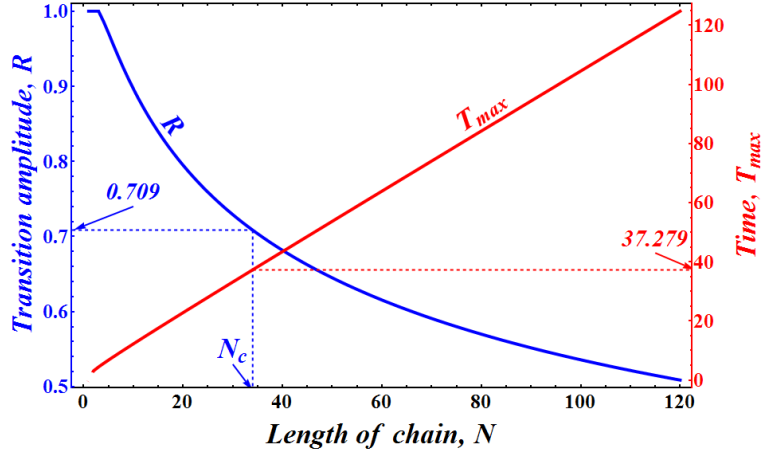,
  scale=0.3
   ,angle=0}
\caption{The maximum   $R$ of the  amplitude $R_N$ (solid decreasing line)
 and the appropriate   time instant $\tau_{max}$  (solid increasing straight line) as  functions of $N$.
 The plateau in the beginning of the decreasing line reflects the fact that $R(2)=R(3)=1$. 
 The similar plateau appears in other figures below.
 The critical length 
 $N_c=34 $ is defined through the critical value  $R_c=\frac{1}{\sqrt{2}}$ (see eq.\ref{conRR}):
 $R(N_c+1)=0.704 <R_c < R(N_c)=0.709 $.  
 This is the maximal chain length allowing us to create the states with 
 zero polarization and non-zero coherence intensity for any value of the inverse temperature $b$, 
 see Sec.\ref{Section:cohNc}.
 The value  $R(N_c)$ and the appropriate value $\tau_{max}(N_c)=37.279$ 
 are indicated in this figure.
} 
  \label{Fig:FT} 
\end{figure*}

Essentially, it is  the function  $R_N(\tau)$
 that is responsible for the creatable region.  In fact, one can verify that the 
creatable region increases with an increase in $R_N$ and the whole receiver's state space 
can be created if $R_N=1$. This situation is realized only for the case $N=2$ and $N=3$ \cite{CDEL}, when
$R_2(\tau) =  \sin\frac{\tau}{2}$ and  $R_3(\tau) =  \sin^2\frac{\tau}{2\sqrt{2}}$, so that 
$R_2(\pi)=1$  and $R_3(\pi \sqrt{2})=1$.
In this case the whole receiver's state space can be created. 
For $N>3$, the transition amplitude is a combination of the  $\tau$-oscillating functions with 
non-rational frequencies resulting in $R_N\to 1$ only over the very
long time interval. 
So, over a reasonable time interval, we always have 
$R_N<1$ and  the creatable region does not cover the
whole receiver's state space. 
This conclusion is justified below in Fig.\ref{Fig:regPolN2to120}. 
But, to maximize the  creatable  region, 
we can pick up the time instant corresponding to  the largest $R_N$. In this regard we notice, that 
the  function $R_N(\tau)$ (at fixed $N>3$) has a well-formed 
decreasing sequence of 
maxima separated by the rather
long time intervals, which  can be approximated by the 
linear combination of the Bessel functions as shown in Appendix B.
Thus, we 
consider the state creation at the time instant $\tau_{max}(N)$ corresponding to the first maximum 
 (the biggest one over the time interval $0<\tau\lesssim 2 N$ in our numerical simulations) 
of the function $R_N(\tau)$. We denote this maximum by $R(N)$ or just by $R$ and rewrite formulas (\ref{IJ0}) and 
(\ref{Jdef0}) as 
\begin{eqnarray}\label{IJ}
&&
I
=
\frac{1}{2}\left((1-R^2)\tanh\frac{{b}}{2}-R^2 \cos (\alpha \pi) \right)
,\\\label{Jdef}
&&
J = 
\frac{1}{4} R^2 \sin^2(\alpha\pi)  \left(\tanh\frac{{b}}{2}\right)^{2(N-1)},
\end{eqnarray}
 which we refer to in describing  the creatable region.
The maximum $R(N)$ and the appropriate 
time instant $\tau_{max}(N)$ are shown in Fig.\ref{Fig:FT} as functions of $N$. We see that 
$\tau_{max}(N)$
is essentially a  linear function of $N$. One should note that the 
points $R(2)$ and $R(3)$ in this figure form a small plateau, 
reflecting  the fact that 
the whole receiver's state space is creatable in the cases 
 $N=2,3$.  
For $N>3$, the maximum $R(N)$ 
is a rapidly decreasing function of $N$. 

\subsection{Map of  control-parameter space into  creatable-parameter region}
\label{Section:CRREG}
\begin{figure*}
   \epsfig{file=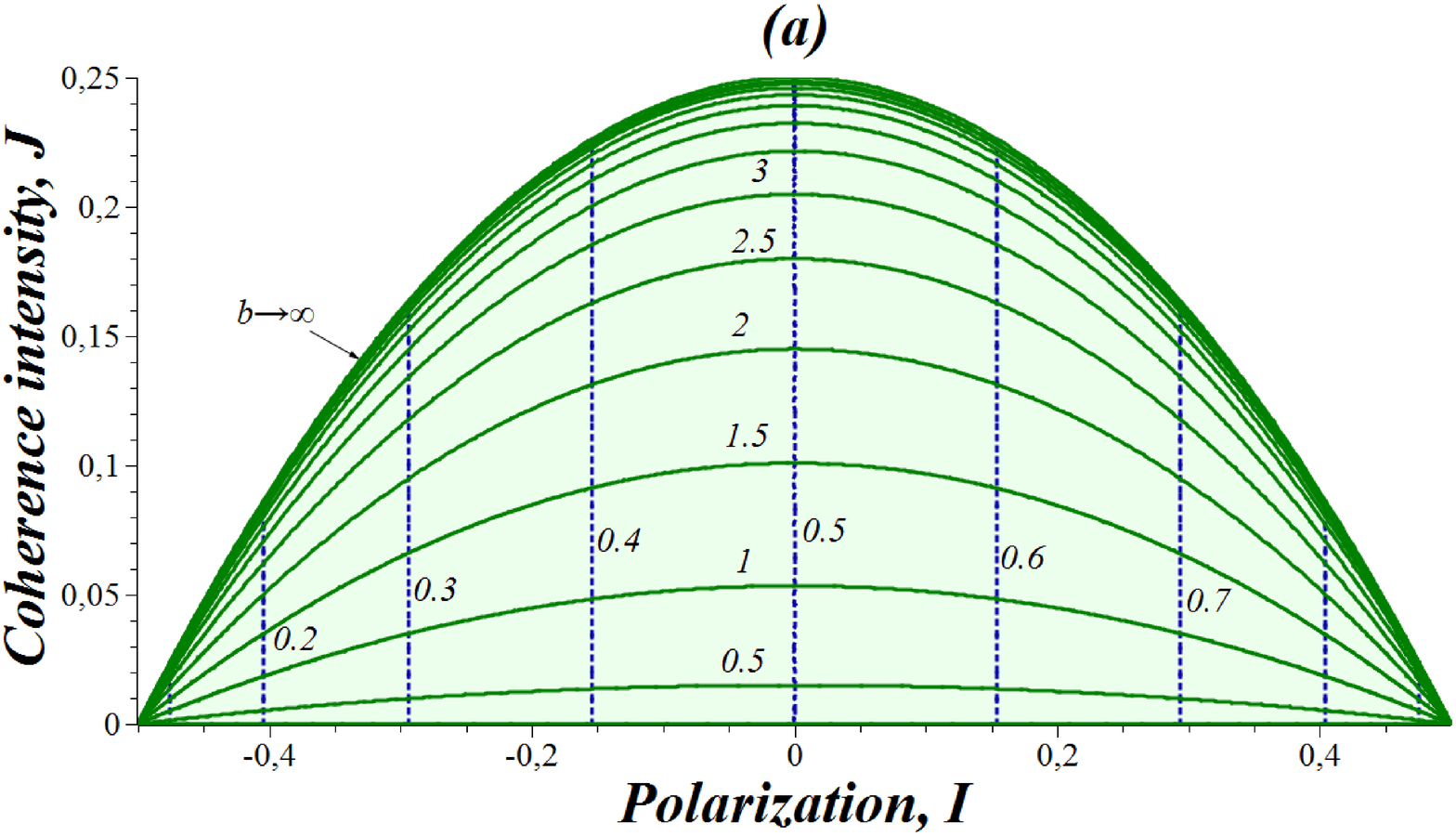,
   %NnPol2,
  scale=0.25
   ,angle=0%270
} \epsfig{file=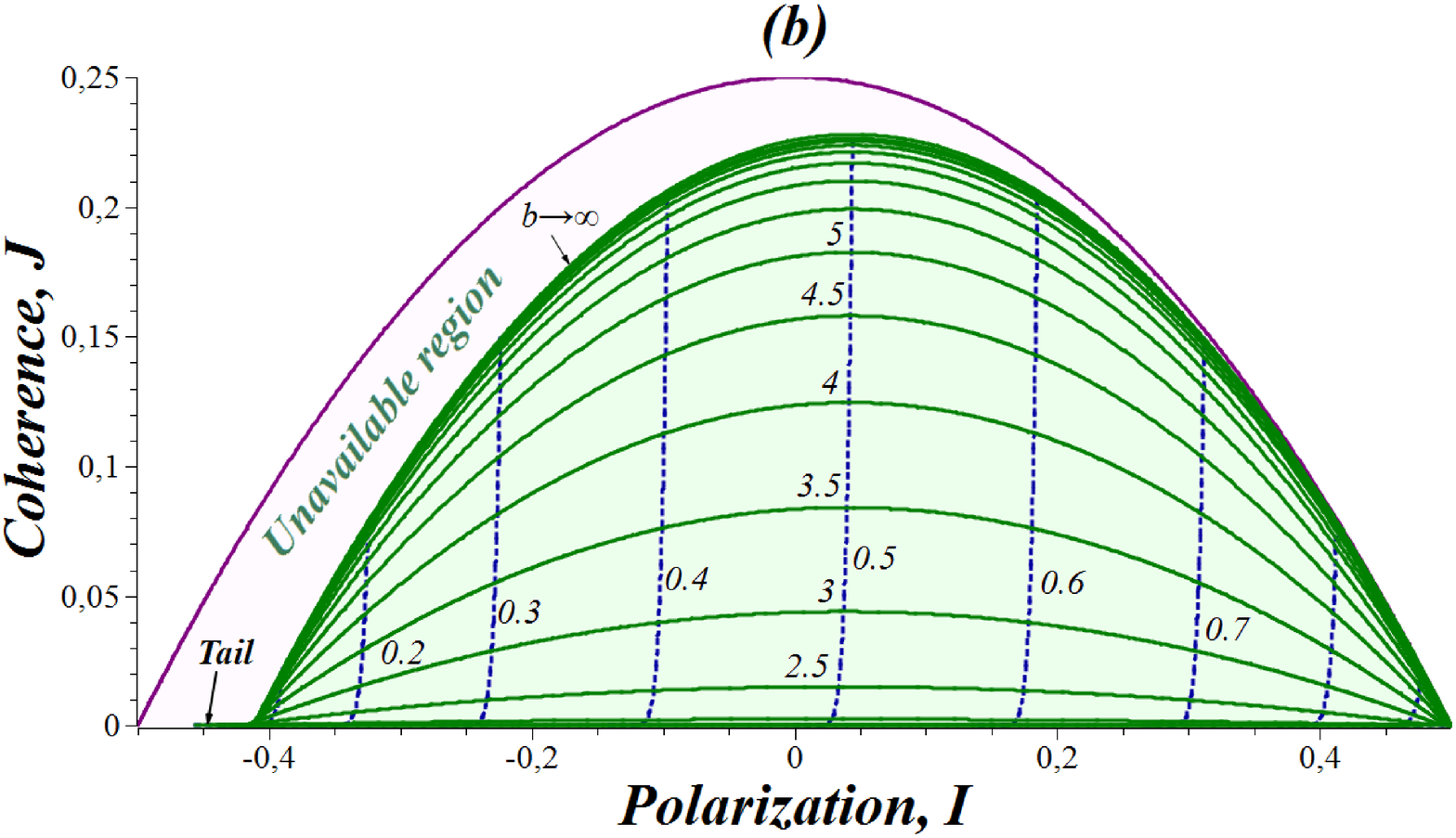,
%NnPol6,
  scale=0.25
   ,angle=0%270
}\newline
\epsfig{file=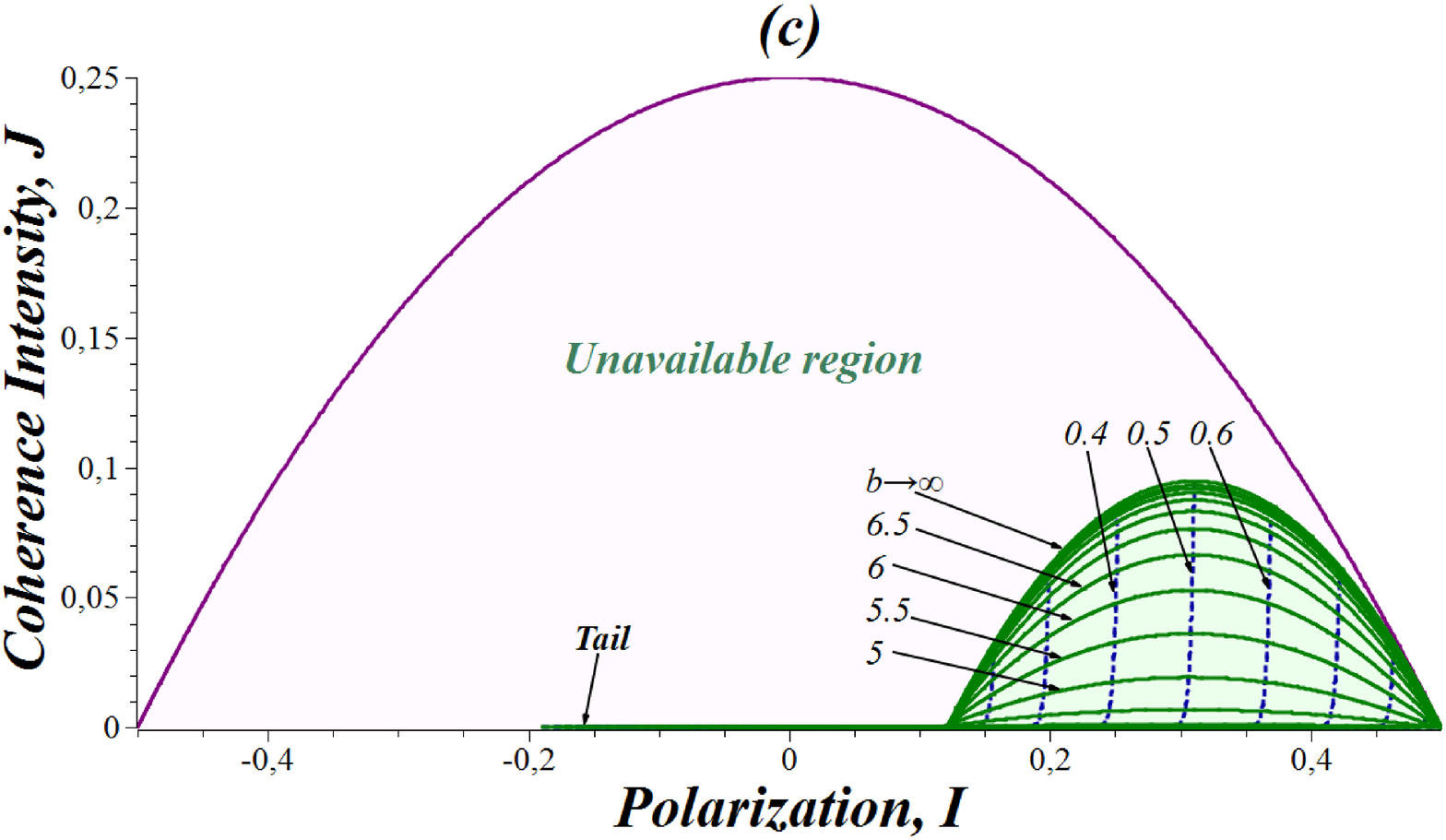,%NnPol60
  scale=0.25
   ,angle=0%270
}
\epsfig{file=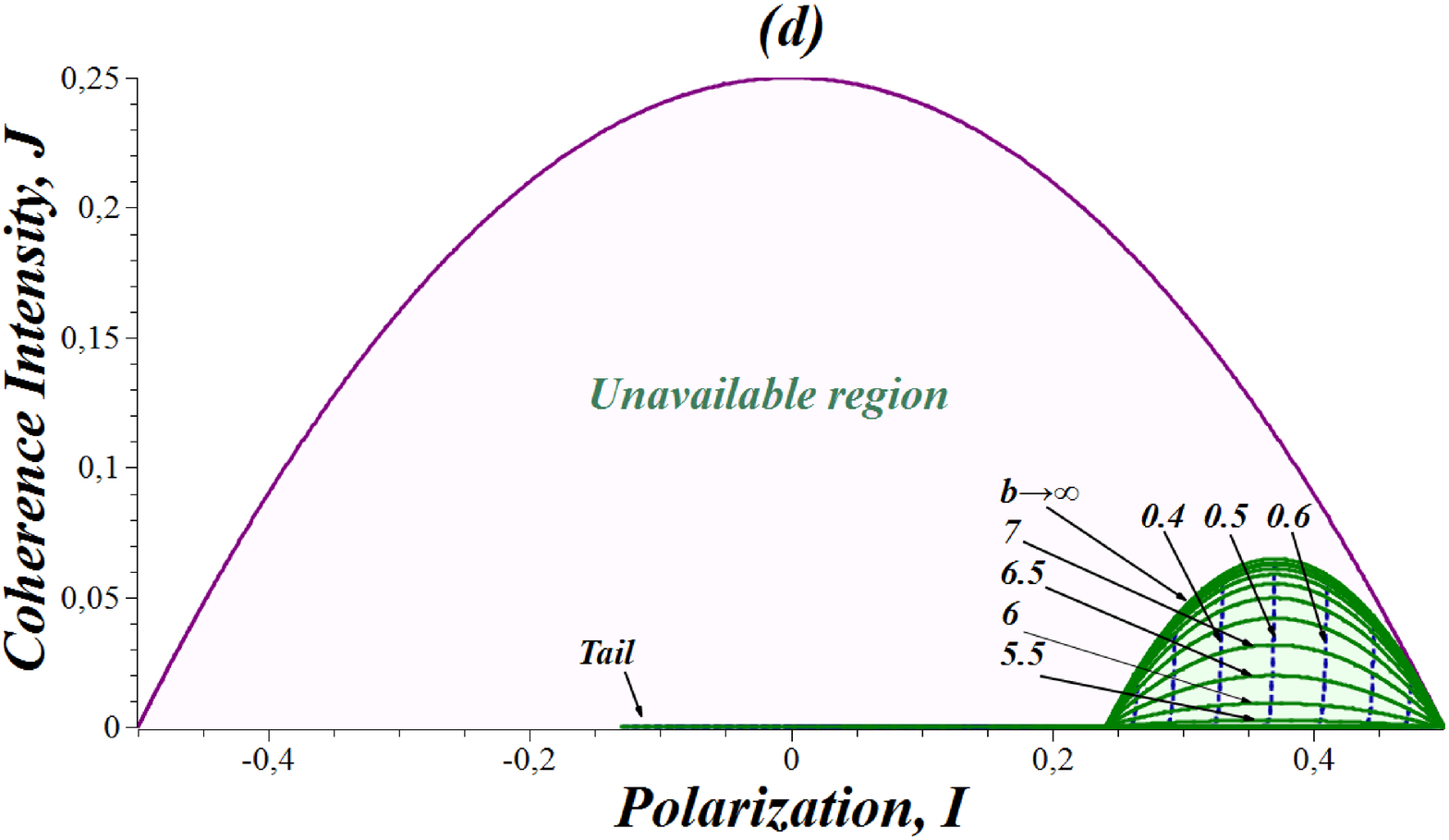,%NnPol120
  scale=0.25
   ,angle=0%270
}
\caption{The creatable regions  in the plane ($J,I$) for the chains of different lengths:
(a) $N=2$, (b) $N=6$, (c) $N=60$ and (d) $N=120$. 
Here, as well as in all the figures below, we use gridding (\ref{gridding}), therewith 
the solid- and dash-lines correspond to, respectively, $b=const$ and $\alpha=const$.
The curves $b=const$ concentrate near the line $b\to\infty$. With an increase in $N$, their 
density near $b=0$ increases as well. 
The region of the receiver's state space between the surrounding solid (violet) line (the parabola representing the boundary of 
the whole receiver state-space) 
and the bell-shaped creatable region is the unavailable region of the receiver state-space, which disappears  for $N=2,3$. 
 The well-formed tail of 
 states 
 with the vanishing coherence intensity is  depicted in the cases $N=6$, 60 and 120 (figs. (b)-(d))
 to the left of the bell-shaped region. In terms of the control parameters, 
 this tail corresponds   to the limit  $\alpha \to 0$ (any $b$).
 The polarization of the tail's end-point  can be  calculated 
 by the formula $I=-\frac{R^2}{2}$, see eq.(\ref{intI}).
 With an increase in $N$, the creatable region  shrinks
 to the point $(I,J)=(0,0)$, while the tail's polarization
 becomes  the interval 
 $0\le I \le \frac{1}{2}$ covering the whole possible  positive interval of the creatable polarization, 
 see Sec.\ref{Section:tail}.}
  \label{Fig:regPolN2to120} 
\end{figure*}

Apparently  \cite{BZ_2015}, the spin dynamics in 
long chains compresses the 
creatable region so that the whole receiver's state-space is divided into two parts: the 
creatable and unavailable regions. 
To visualize the creatable region, we 
represent the map
\begin{eqnarray}
\label{abIJ}
 (\alpha,{b}) \stackrel{(\ref{IJ},\ref{Jdef})}{\to} (I,J)
 \end{eqnarray}
 (which is explicitly given by formulas (\ref{IJ}) and (\ref{Jdef}))
 in Fig.\ref{Fig:regPolN2to120}
 for the chains of different lengths $N$. As was mentioned in Sec.\ref{Section:RN}, 
 the whole state space of the one-qubit receiver can be created using  the  chain of two  (or three) nodes, Fig.\ref{Fig:regPolN2to120}a. 
 If $N>3$, then the unavailable region appears, which is shown in  Figs.\ref{Fig:regPolN2to120}(b-d)
  (where the surrounding solid line is 
 the  boundary of the 
whole receiver's state space  (\ref{boundary})). In Fig.\ref{Fig:regPolN2to120}, the dash-lines correspond to
 the constant values of the parameter $\alpha$, while the 
 solid lines correspond to the 
constant values of the parameter ${b}$, therewith we use the following gridding throughout this paper:
 \begin{eqnarray}\label{gridding}
 \begin{array}{ll}
 \alpha= 0.1 n,\;\;n=0,1,\dots,10 & {\mbox{dash-lines}},\cr
{b}=0,\;0.1, \;\;0.5 n,\;\;n=1,2,3,\dots  &{\mbox{solid lines}}.
\end{array} 
 \end{eqnarray}
 We see that the solid lines nestle to the  either line 
 ${b} \to \infty$ or  line ${b} = 0$, which is most clearly shown in 
 Fig.\ref{Fig:regPolN2to120}c,d where only a few solid lines are well separated from the top and from the bottom of 
 the bell-shaped creatable region. 
 The shrinking to the line ${b}=0$  means that, increasing the  chain length, we must  decrease the 
 temperature (or increase $b$) to create the valuable 
 coherence intensity.
 As $N\to \infty$, 
 the creatable region compresses to the right corner of the receiver's state space, which, consequently, 
 is the most reproducible  area of the state space. On the contrary, the 
 states in the left corners of the 
 receiver's state space become unavailable already in short chains, see Fig.\ref{Fig:regPolN2to120}b.

The common feature of the creatable regions in Fig.\ref{Fig:regPolN2to120} is the tail (vanishing in 
chains of length $N=2,3$) of states 
with negligible coherence intensity situated to the left of the bell-shaped creatable region.
This tail represents the   ''classical'' part of polarization, i.e., 
such polarization that causes negligible quantum effects described by the coherence intensity. 

Notice, although  the creatable region is mainly the one-to-one image of  map 
(\ref{abIJ}), there is a small  subregion which is the  two-fold image of map (\ref{abIJ}). 
This feature is   discussed in Sec.\ref{Section:boundary}.

 \subsection{One-to-one and two-to-one mapped creatable sub-regions}
 \label{Section:boundary}
As mentioned in the Introduction, the whole creatable region is divided into the two 
subregions. One of them is the large subregion under the  curve $b\to\infty$ in the plane $(I,J)$.
This is the  one-to-one image of map (\ref{abIJ}) (there is no intersections of solid lines inside of this region in
Fig.\ref{Fig:regPolN2to120}) 
with the control parameters inside of the 
following sub-domain:
\begin{eqnarray}\label{OneToOne}
&&
\alpha^{br}(b) < \alpha\le 1  ,\;\;0\le b\le\infty\;\; \;\;{\mbox{sub-domain of the one-to-one map (\ref{abIJ})}}.
\end{eqnarray}
Another subregion is the small area 
above the curve $b\to\infty$ (including the tail).
This is the two-fold image of map  
(\ref{abIJ}) (in other words, the function $J(I)$ is a two-sheet function in this subregion) 
with the control parameters inside of the other
  sub-domain:
\begin{eqnarray}
\label{TwoFold}
&&
0\le\alpha\le \alpha^{br}(b) ,\;\;0\le b\le\infty\;\; \;\;{\mbox{sub-domain of the two-to-one map (\ref{abIJ})}},
\end{eqnarray}
Any point $(I,J)$ from this subregion can be created using  two different pairs of the control 
parameters $(\alpha,b)$. This means, in particular, that two different values of  temperature (parameter $b$) 
can create the same polarization and 
coherence intensity if the parameter $\alpha$ of the  sender's pure initial state properly depends on $b$.
The boundary line $\alpha^{br}(b)$  in the plane of control parameters (separating the  two above sub-domains 
(\ref{OneToOne}) and (\ref{TwoFold}))  is defined  below in eq.(\ref{alpbr}).

\subsubsection{One-to-one mapped subregion}
\label{Section:onetoone}
First of all we observe that there is a curve covering the top of the
bell-shaped creatable region, see Fig.\ref{Fig:regPolN2to120}. It corresponds to the
limit 
${b}\to \infty$:
\begin{eqnarray}\label{liminfI}
&&
I_{\infty}=\frac{1}{2}\Big((1-R^2) - R^2\cos(\alpha \pi)\Big),\\\label{liminfJ}
&&
J_{\infty}=\frac{R^2}{4} \sin^2(\alpha\pi),\;\;0\le \alpha\le 1.
\end{eqnarray}
Formula (\ref{liminfJ}) gives us the maximal value of $J$, $J^{max}_{\infty}$, at $\alpha=\frac{1}{2}$:
\begin{eqnarray}\label{Jmax}
J^{max}_{\infty}=\frac{R^2}{4}, 
\end{eqnarray}
therewith, the appropriate polarization $I_{J_{max}}$ reads
\begin{eqnarray}\label{Imax}
I_{J_{max}}=\frac{1-R^2}{2}\ge 0.
\end{eqnarray}
In particular, for $N=2$ and 3,  we have $R(2)=R(3)=1$ and $J^{max}_{\infty} =\frac{1}{4}$, $I_{J_{max}}=0$.
Notice that $J_{\infty}$ vanishes at $\alpha=0,1$, therewith 
\begin{eqnarray}\label{Ic}
I_{\infty}|_{\alpha=0}\equiv I_c =\frac{1}{2}-R^2, \;\;I_{\infty}|_{\alpha=1}=\frac{1}{2}.
\end{eqnarray}
We must emphasize that the parametrically represented  curve (\ref{liminfI},\ref{liminfJ})  is the upper boundary of the 
image of   one-to-one map (\ref{abIJ},\ref{OneToOne}).

\subsubsection{Creatable two-to-one mapped  subregion}
\label{Section:twofold}
\begin{figure*}
\epsfig{
file=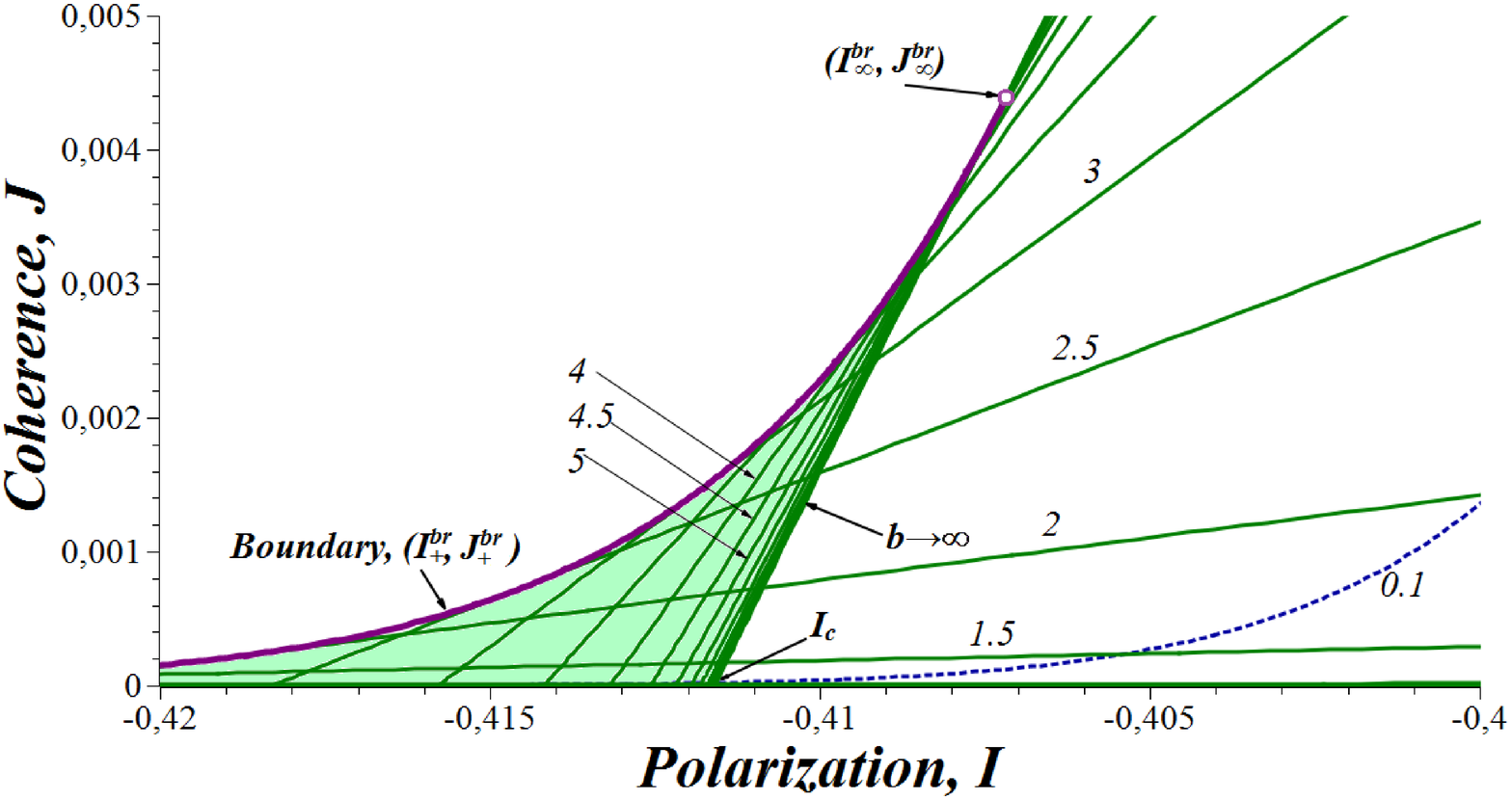,
  scale=0.35
   ,angle=0%270
}
\caption{The left
corner of the bell-shaped creatable region   for the chain of $N=6$ nodes.
The shaded area above the curve $b\to\infty$ (the boundary of the one-to-one mapped region)
is covered by the two-to-one map
 (\ref{abIJ},\ref{TwoFold}), except for  the upper boundary (one-to-one mapped),
 described by the points $(I^{br}_+,J^{br}_+)$ given in formulas 
 (\ref{Jextr}) and (\ref{Iextr}) (the upper  bold-solid  (violet) line in the figure). 
 The branch point $(I^{br}_\infty,
J^{br}_\infty) = (-0.407,0.004)$ is the cross-point of this boundary line with the boundary of 
the one-to-one mapped region. The left point of the boundary line, corresponding  to $b=0$, is
$(-\frac{R^2}{2},0)
\Big|_{N=6}=(-0.456,0)$ (this point is not shown in the graph).
Therewith, the creatable area to the left of  
  $I_c=-0.412$, $I<I_c$, 
  corresponds to the two-sheet function $J(I)$ defined for all  creatable values of $J$. 
  Moreover, 
  this subregion is  creatable  only at finite $b$ (non-zero temperature). 
We use gridding (\ref{gridding}), so that 
only one dash-line $\alpha = 0.1$ 
appears in  this figure.
}
  \label{Fig:nonun} 
\end{figure*}

The two-to-one mapped subregion 
is a  new feature of the remote state-creation process, which was not investigated  before 
and thus deserves a special 
consideration. 
This subregion includes the 
tail to the left of the bell-shaped region and the close neighborhood of the left corner of
this region in Fig.\ref{Fig:regPolN2to120}. 
Since the scale of  Fig.\ref{Fig:regPolN2to120} does not allow us  to observe
 this subregion, we represent this subregion  for the case  $N=6$ using a suitable scale
in Fig.\ref{Fig:nonun}.  Notice that the two-to-one mapped 
subregion disappears if $R=1$, i.e., $N=2,3$. 

Now we derive the boundaries of the two-to-one mapped subregion.
Apparently, 
 map (\ref{abIJ}) is the one-to-one map at  a point $(I,J)$  if $J|_{I=const}$ 
is a monotonic function of ${b}$  at this point. In the case of the  two-to-one mapping, $J|_{I=const}$  
as a function 
of $b$ loses its monotonic behavior and acquires the extremal point (a maximum).  
 The appearance of such a maximum can be used
as an indicator of the two-to-one mapping. 

The upper boundary of this two-to-one mapped subregion can be
parametrically described as follows. 
First, we   
solve eq.(\ref{IJ}) for  $\alpha$,
\begin{eqnarray}\label{alpIJ}
\cos(\alpha\pi) = \frac{1}{R^2}\Big( (1-R^2)\tanh\frac{b}{2} - 2 I\Big),
\end{eqnarray}
and substitute $\alpha$ from eq.(\ref{alpIJ})
into eq.(\ref{Jdef}) obtaining $J$ as a function of parameters 
$I$ and $b$, which now are considered as independent ones:
\begin{eqnarray}\label{J2}
J(I,b)=\frac{\left(\tanh\frac{{b}}{2}\right)^{2(N-1)}}{4 R^2}\left(
R^4-\Big(2I -(1-R^2) \tanh\frac{{b}}{2}\Big)^2
\right).
\end{eqnarray}
Next, we calculate  the 
 partial derivative of $J$ 
with respect to $b$  and equate it to zero  (we cancel positive factors and 
use the superscript ''$br$''  to mark quantities associated with the  
 two-sheet subregion):
\begin{eqnarray}\label{pol2I2}
&&
4 (I^{br})^2(N-1) - 2 I^{br}\tanh\frac{b^{br}}{2} (2 N-1) (1-R^2) +\\\nonumber
&&
\tanh^2\frac{b^{br}}{2} N (1-R^2)^2 -R^2(N-1) =0.
\end{eqnarray}
The lhs of eq.(\ref{pol2I2}) is a  quadratic expression
in both $I^{br}$ and $\tanh\frac{b^{br}}{2}$. Let us solve it 
for the polarization $I^{br}$:
\begin{eqnarray}\label{Iextr}
I^{br}_{\pm} = \frac{1}{4(N-1)}\left(
(2 N-1) (1-R^2) \tanh\frac{{b^{br}}}{2} \mp \sqrt{4 (N-1)^2 R^4 + 
(1-R^2)^2 \tanh^2\frac{{b^{br}}}{2}}\right).
\end{eqnarray}
Substituting this  polarization into eq.(\ref{J2}) we obtain the following expression for the coherence intensity:
\begin{eqnarray}\label{Jextr}
J^{br}_\pm=\frac{(1-R^2)R^2\left(\tanh\frac{{b^{br}}}{2}\right)^{2N-1}}{
2\Big((1-R^2) \tanh\frac{{b^{br}}}{2} \pm \sqrt{4 (N-1)^2 R^4 + 
(1-R^2)^2 \tanh^2\frac{{b^{br}}}{2}}\Big)}.
\end{eqnarray}
However, one can see that $J^{br}_-$ in  formula (\ref{Jextr}) is
negative, which is physically impossible.
Thus, there is only one extremum point $(I^{br}_+,J^{br}_+)$ which is the maximum. 
The right hand sides  of eqs.(\ref{Iextr}) and (\ref{Jextr}) are defined for all $N$ and $b^{br}$, 
i.e.,  any pair of values of these parameters uniquely defines the
polarization $I^{br}_{+}$ such that  the  coherence intensity $J(I^{br}_+,b)$ given in eq.(\ref{J2}) increases with $b$ till 
$b=b^{br}$, where $J$ takes its maximal value $J^{br}_+$, and then decreases either  to its limit value 
$J(I^{br}_+,\infty)>0$  or to zero. 
Thus, for a given $I^{br}_+$, the  interval of 
coherence intensity 
$\max\Big(0,J(I^{br}_+,\infty)\Big) \le J < J^{br}_+$ is twice covered by  map (\ref{abIJ}).
Therewith, 
\begin{eqnarray}\label{Ibrc}
0\le J < J^{br}_+, \;\;{\mbox{if}}\;\; I^{br}_+ \le I_c ,
\end{eqnarray}
where $I_c$ is defined  in (\ref{Ic}) as a polarization corresponding to $J_{\infty}=0$.
In addition, the states  with $I< I_c$ 
can be created   only at finite $b$
(finite temperature) and can  not exists as $b\to\infty$ (zero temperature). The polarization $I_c$ is marked in Fig.\ref{Fig:nonun} for $N=6$.

To estimate the size of the two-to-one mapped subregion, we calculate 
the maximal value of $J^{br}_+$ for a given $N$. It is clear that 
this maximum corresponds to  ${b}\to\infty$:
\begin{eqnarray}
J^{br}_{\infty}=\frac{(1-R^2)R^2}{
2\Big((1-R^2)  + \sqrt{4 (N-1)^2 R^4 + (1-R^2)^2 }\Big)}.
\end{eqnarray}
The appropriate expression for   $I^{br}_{\infty}$ reads:
\begin{eqnarray}\label{Iextr2}
I^{br}_{\infty} = \frac{1}{4(N-1)}\left(
(2 N-1) (1-R^2) - \sqrt{4 (N-1)^2 R^4 + (1-R^2)^2 }\right).
\end{eqnarray}
Thus, the upper boundary is parametrically defined by eqs.(\ref{Iextr}) and  (\ref{Jextr}) 
(with the parameter 
$b^{br}$, $0 \le b^{br} < \infty$) over the interval
\begin{eqnarray}\label{intF2fold1}
-\frac{R^2}{2}\le I\le I^{br}_\infty.
\end{eqnarray}
This boundary for the 6-node chain  
is  represented in Fig.\ref{Fig:nonun} by 
the bold-solid (violet) line.  
Notice that $I^{br}_{\infty}$ changes its negative sign to the positive one in passing from 
$N=33$ to $N=N_c=34$.
The point 
$(J^{br}_\infty,I^{br}_\infty)$ can be referred to as a branch point because, 
passing through this point, the 
function $J(I,b)$ becomes a two-sheet function of $b$. This point is marked in Fig.\ref{Fig:nonun} ($N=6$).
Notice also that, by construction, the upper boundary itself is ones covered by the control parameters.

%The appropriate boundary $\alpha^{br}$ in the space of control parameters $(\alpha,b)$ 
%can be derived substituting $I^{br}_+$ from eq.(\ref{Iextr})  into
%eq.(\ref{alpIJ}):
%\begin{eqnarray}\label{alpbr}
%&&
%\cos(\alpha^{br}(b)\pi)=
%\\\nonumber
%&&\frac{1}{2 R^2 (N-1)}\left(
%\sqrt{4 R^4( N -1)^2 +(1-R^2)^2 \tanh^2\frac{b^{br}}{2} } - (1-R^2)\tanh\frac{b^{br}}{2}\right),
%\end{eqnarray}
Regarding the right boundary of the two-to-one mapped subregion (see the right boundary of the shaded area in 
Fig.\ref{Fig:nonun}), it is described by the part of the curve $b\to\infty$ 
(\ref{liminfI},\ref{liminfJ}) over the interval
\begin{eqnarray}
\label{intF2fold2}
I_c\le I\le I^{br}_\infty.
\end{eqnarray}
This boundary is an image of the   boundary $\alpha^{br}(b)$ in the control parameter 
plane (separating the preimages of the one-to-one and 
two-to-one mapped subregionÙ), which
was used in eqs.(\ref{OneToOne},\ref{TwoFold}). 
This boundary consists  of the intersection points  of the curves $b=const$ with the curve $b\to\infty$, i.e, 
 $\alpha^{br}$ satisfies the following system
\begin{eqnarray}\label{IJinf}
I(\alpha^{br},b) = I(\tilde\alpha,\infty),\\\nonumber
J(\alpha^{br},b) = J(\tilde\alpha,\infty)
\end{eqnarray}
for some $\tilde\alpha$.
Here, $I(\alpha,b)$ and $J(\alpha,b)$ are given in eqs.(\ref{IJ}) and (\ref{Jdef}).
Solving this system we obtain
\begin{eqnarray}\label{alpbr}
\cos(\alpha^{br}\pi)&=&\frac{(1-R^2) (1-\tanh\frac{b}{2}) \tanh^2\frac{b}{2}-\sqrt{D}}{R^2\left(
\left(\tanh\frac{b}{2}\right)^{2N}-\tanh^2\frac{b}{2}\right)},\\\nonumber
D&=&R^4 \Big(\tanh^4\frac{b}{2} + 
\tanh^{4 N}\frac{b}{2}\Big) +\\\nonumber
&&
 \left(\tanh\frac{b}{2}\right)^{2(N+1)} \Big(\Big(1-\tanh\frac{b}{2}\Big)^2 (1- 2 R^2) +R^4 (\tanh^2\frac{b}{2} -
2 \tanh\frac{b}{2} -1)\Big).
\end{eqnarray}
%This boundary  divides the whole control-parameter space into mentioned above
%subregions (\ref{OneToOne}) and (\ref{TwoFold}), 
%which, respectively,   ones and  twice cover the appropriate creatable 
%subregion of parameters $I,J$ through  map  (\ref{abIJ}).

%Finally, eq.(\ref{alpbr}) yields us an appropriate expression for $\alpha^{br}_\infty =\alpha^{br}|_{b\to\infty} $:
%\begin{eqnarray}\label{alpbrinf}
%\cos(\alpha^{br}_\infty\pi)=\frac{1}{2 R^2 (N-1)}\left(
%\sqrt{4 R^4( N -1)^2 +(1-R^2)^2 } + R^2-1\right).
%\end{eqnarray}

%%%%%%%%%%
\subsubsection{Interval of creatable polarization and tail with vanishing coherence intensity}
\label{Section:tail}
The interval of creatable polarization at a given $b$ is defined by the boundary values of 
the parameter $\alpha$. Thus, at $\alpha=0$ and $\alpha=1$,  we obtain, respectively, 
the lower and upper  boundaries of polarization as functions of $b$ from eq.(\ref{IJ}):
\begin{eqnarray}\label{Ilow}
I_{low}(b)=\frac{1}{2} \left((1-R^2)\tanh\frac{b}{2}- R^2\right)\\\label{Iup}\
I_{up}(b)=\frac{1}{2} \left((1-R^2)\tanh\frac{b}{2}+ R^2\right).
\end{eqnarray}
Therewith the minimal value of $I_{low}$ corresponds to $b=0$, while the maximal 
value of $I_{up}$ corresponds to $b\to\infty$; thus the 
maximal variation interval of $I$ is the  following one:
\begin{eqnarray}\label{intI}
-\frac{R^2}{2}= I_{low}(0)\le I \le I_{up}(\infty)=\frac{1}{2}
\end{eqnarray}
In the case $N=2,3$, we have  $R\equiv 1$, so that inequality  (\ref{intI}) reaches   the
boundary of the receiver's state space: $|I|\le \frac{1}{2}$. 

Considering the polarization at the fixed temperature,
$I_{low}(b)\le I \le I_{up}(b)$, we point out the two following limit intervals:
\begin{eqnarray}\label{binf}
&&
\frac{1}{2}(1-2 R^2) \le I \le \frac{1}{2} , \;\;\;b\to \infty\\\label{b0}
&&
-\frac{R^2}{2} \le I \le \frac{R^2}{2} ,\;\;\;b=0.
\end{eqnarray}

As $N\to \infty$  (consequently, $R\to 0$), each of the above  intervals shrinks into a point, 
and formulas (\ref{IJ}) and (\ref{Jdef}) show that the  creatable polarization is completely  
defined by the temperature (the parameter $\alpha$ of the  sender's  pure initial 
state disappears in this limit): $I\to \frac{1}{2}\tanh\frac{b}{2}$, 
and $J \to  0$. Thus, the variation interval of  the polarization tends to
$0 \le I \le \frac{1}{2}$ covering the whole positive interval of the creatable polarization, while
the coherence intensity tends to zero for any temperature. This is a principal difference 
between the polarization and the coherence intensity.
Notice that the above  limit interval  coincides with the limit $N\to \infty$ of the tail of polarization
(see Fig.\ref{Fig:regPolN2to120}), which  covers the interval
 $-\frac{R^2}{2} \le I \le I^{br}_\infty$ in the case of  finite $N$.

\subsection{Mutual relations between creatable polarization and coherence intensity}
Although the parameters of the creatable region discussed in Sec.\ref{Section:boundary} 
can be referred to as characteristics of the creatable region, they  describe mainly the boundary of this region. 
Now we introduce some characteristics reflecting mutual relation between creatable 
polarization and coherence intensity inside of the creatable region. 

\subsubsection{States with  zero polarization}
\label{Section:cohNc}
\begin{figure*}
\epsfig{file=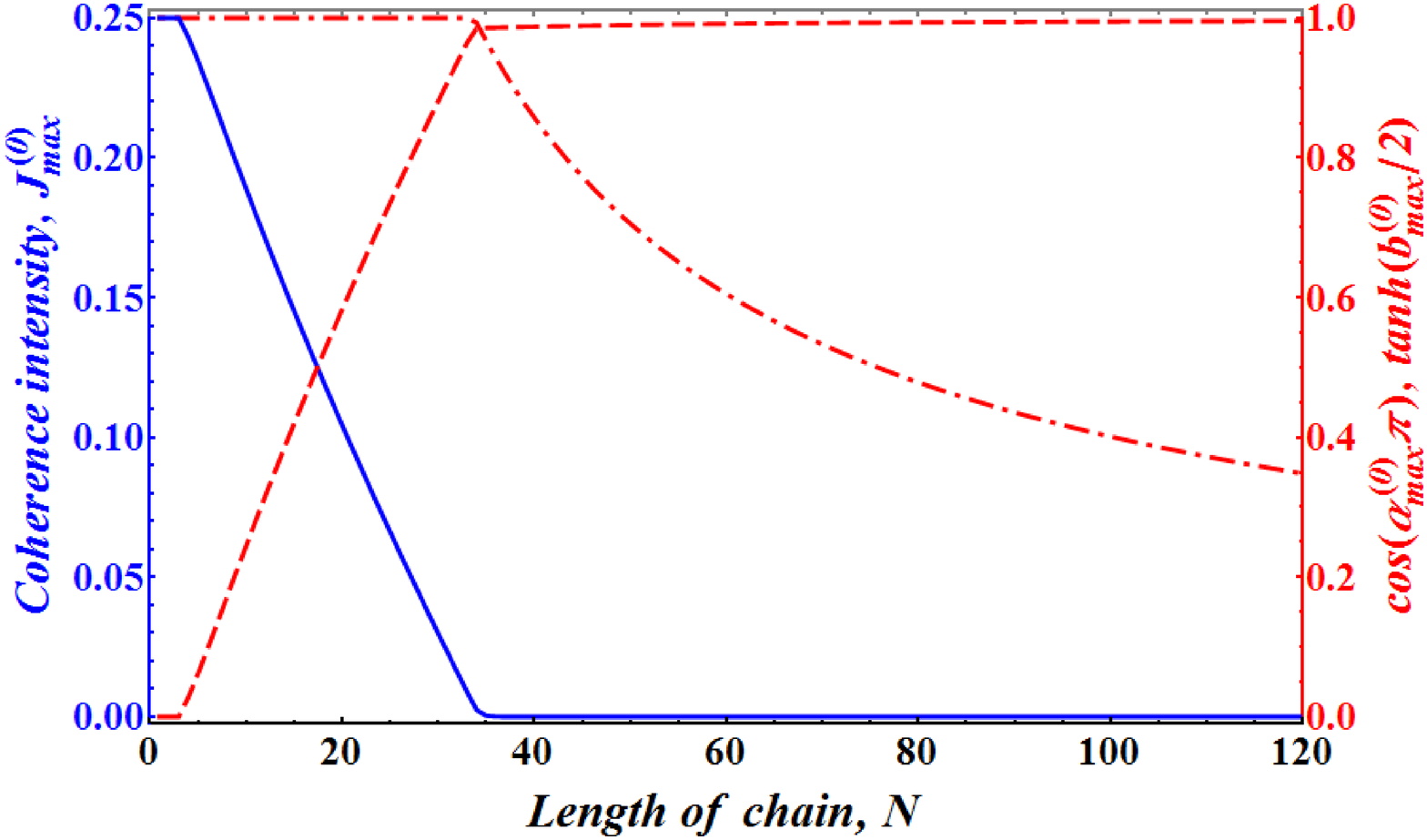,
  scale=0.3
   ,angle=0%270
}
\caption{
The characteristics of the states with zero polarization. 
The creatable coherence intensity  $J^{(0)}_{max}$ (solid line)  and the appropriate control parameters 
(in the form  of  
 $\tanh\frac{{b}^{(0)}_{max}}{2}$ (dash-dot-line) and 
$\cos(\alpha^{(0)}_{max}\pi)$ (dash-line))  are given 
as functions of  $N$. All graphs have breakpoints at $N=N_c=34$. 
For $N>N_c$, the  coherence intensity vanishes very rapidly with an increase in $N$, see (\ref{J0v}).}
  \label{Fig:I0} 
\end{figure*}

In this subsection, we consider the case $I=0$, when the
coherence intensity (which is associated with quantum effects)
is created without polarization 
(classical effect). In this case, eq.(\ref{IJ0}) yields  $\rho_{11}=\frac{1}{2}$, and eq.(\ref{IJ}) results 
in the following 
 relation between the 
control parameters $\alpha$ and  ${b}$  
(we use the superscript  ''$(0)$'' to differ the quantities of this subsection from those 
of the general position):
\begin{eqnarray}\label{cr1}
\cos(\alpha^{(0)} \pi)=\frac{1-R^2}{R^2}\tanh\frac{{b}^{(0)}}{2}>0.
\end{eqnarray}
This relation holds for  any $b^{(0)}$  if 
\begin{eqnarray}\label{conRR}
 \frac{1-R^2}{R^2}\le 1, \;\;\Leftrightarrow \;\;R^2\ge \frac{1}{2}
 \stackrel{{\mbox{def}}}{=} R_c^2.
 \end{eqnarray}
 The direct calculation shows (see also Fig.\ref{Fig:FT}) that
 \begin{eqnarray}
 R(35)=0.704 <R_c < R(34)=0.709.
 \end{eqnarray}
 Consequently, 
 inequalities (\ref{conRR}) hold  
 if
 \begin{eqnarray}
 N\le N_c=34.
\end{eqnarray}
Otherwise, if $N>N_c$, then  eq.(\ref{cr1}) yields the following constraint for   $b^{(0)}$:
\begin{eqnarray}\label{constrIJ}
\tanh\frac{{b}^{(0)}}{2} \le \frac{R^2}{1-R^2}.
\end{eqnarray}
Substituting eq.(\ref{cr1}) into eq.(\ref{Jdef}) we obtain 
\begin{eqnarray}\label{J0}
J^{(0)} = 
\frac{\Big(\tanh\frac{{b}^{(0)}}{2}\Big)^{2(N-1)}}{4R^2}\left(
R^4 - (1-R^2)^2\tanh^2\frac{{b}^{(0)}}{2}
\right), && \;\; \forall \; b^{(0)}, \;\;N\le N_c \;\;{\mbox{or}}
\\\nonumber
&&
{\mbox{for $b^{(0)}$ satisfying (\ref{constrIJ})}},\;\;N>N_c. 
\end{eqnarray}
Considering $J^{(0)}$ in (\ref{J0}) as a function of ${b}^{(0)}$, we find its  maximum 
at
\begin{eqnarray}\label{tanh2}
\tanh\frac{{b}^{(0)}_{max}}{2} =\left\{\begin{array}{ll}
1, & N< N_c\cr
\frac{R^2}{1-R^2}\sqrt{\frac{N-1}{N}},& N\ge N_c\cr
\end{array}\right..
\end{eqnarray}
In deriving eq.(\ref{tanh2}),  we take into account that $\left|\tanh\frac{{b}^{(0)}_{max}}{2} \right|\le 1$ and 
\begin{eqnarray}
\left.
\frac{R^2}{1-R^2}\sqrt{\frac{N-1}{N}} 
\right|_{N=34}=0.993 < 1 < \left.
\frac{R^2}{1-R^2}\sqrt{\frac{N-1}{N}} 
\right|_{N=33}=1.021.
\end{eqnarray}
Thus, substituting eq.(\ref{tanh2}) into eq.(\ref{J0}), we obtain
\begin{eqnarray}\label{J02}
J^{(0)}_{max}=\left\{\begin{array}{cc}
\frac{2 R^2-1}{4R^2}, &N< N_c \cr
\frac{(N-1)^{N-1}R^{2(2N-1)}}{4 N^N (1-R^2)^{2(N-1)}}, &N\ge N_c \cr
\end{array}
\right. .
\end{eqnarray}
In addition, substituting eq.(\ref{tanh2}) into eq.(\ref{cr1}), we have
\begin{eqnarray}\label{alpha02}
\cos(\alpha^{(0)}_{max}\pi)=\left\{\begin{array}{cc}
\frac{1-R^2}{R^2}, &N< N_c =34\cr
\sqrt{\frac{N-1}{N}}, &N \ge N_c 
\end{array}
\right. .
\end{eqnarray}
The dependence of $J^{(0)}_{max}$,  ${b}^{(0)}_{max}$  and  
$\alpha^{(0)}_{max}$   on $N$ is depicted 
in Fig.\ref{Fig:I0}. All graphs have breakpoints at $N=N_c$.
%therewith $J^{(0)}_{max}(33)=8.894 \;\;10^{-3}$ and
%$J^{(0)}_{max}(34)=2.367\;\; 10^{-3}$.
Fig.\ref{Fig:I0} demonstrates that, for $N\ge N_c$, the parameter $J^{(0)}_{max}$ vanishes very rapidly with increase in  $N$. 
Thus, 
\begin{eqnarray}\label{J0v}
J^{(0)}_{max}|_{N=34}=2.367\;\;10^{-3},\;\;J^{(0)}_{max}|_{N=40}=1.782\;\;10^{-8},\;\;
J^{(0)}_{max}|_{N=50}=3.012\;\;10^{-18}.
\end{eqnarray}
This means that 
the significant value of the coherence intensity can not be created without the supplementing polarization in long chains. 
The minimal  polarization   needed for creation of the measurable coherence intensity 
is studied  in the next subsection.

\subsubsection{States with detectable  coherence intensity}
\label{Section:largepol}
Since the   coherence intensity can not be created without the polarization in long chains, 
important characteristics of the state creation are such
minimal  $I^{(1)}_{-}$ and maximal $I^{(1)}_{+}$  values 
of the polarization $I$ (at a fixed ${b}$) that 
the creatable  coherence intensity $J$ is 
valuable (i.e., exceeds some conventional value $J_{min}$)  
inside of the interval $I^{(1)}_{-} \le I \le I^{(1)}_{+}$.

\begin{figure*}
\epsfig{file=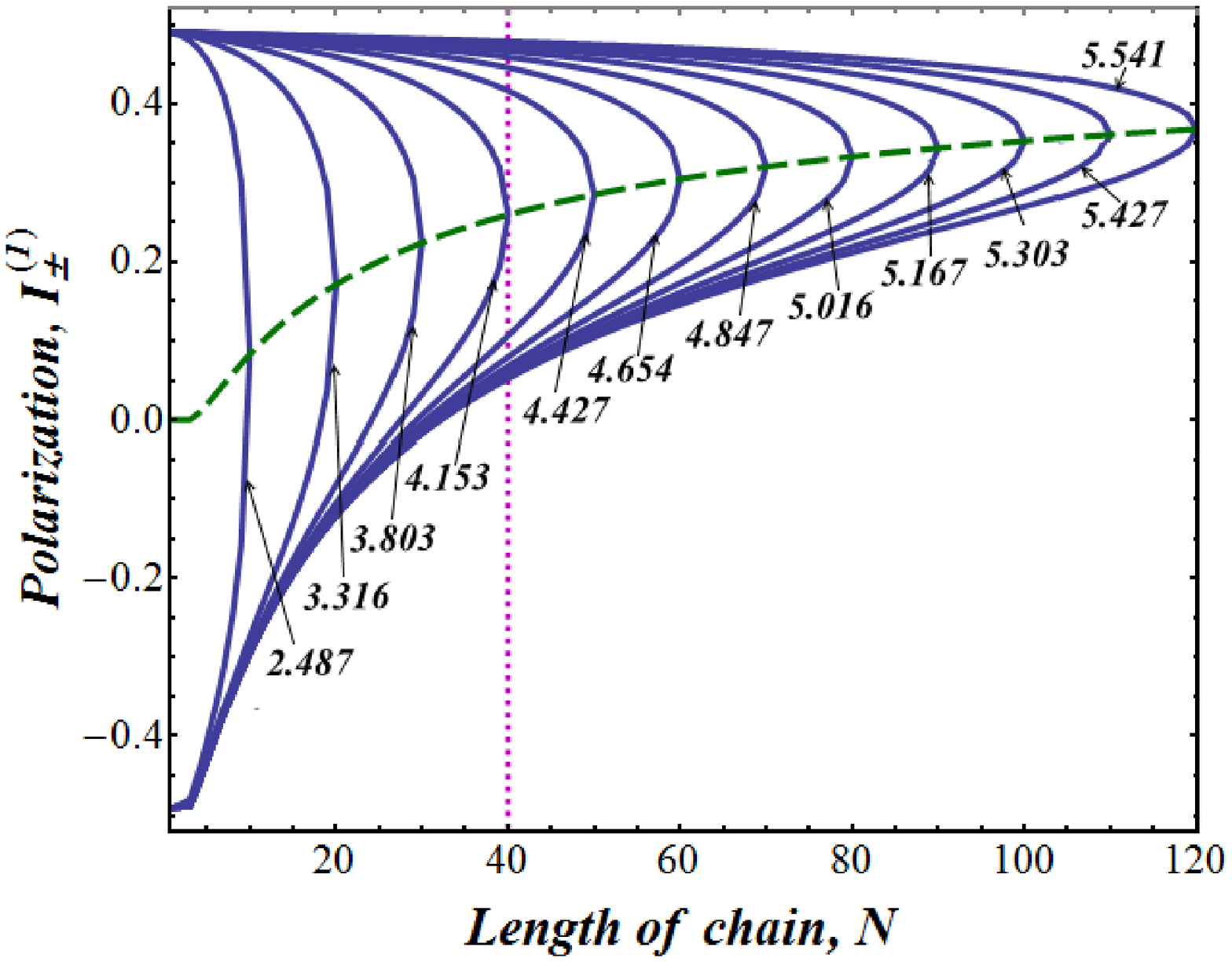,
  scale=0.43
   ,angle=0%270
   }
\epsfig{file=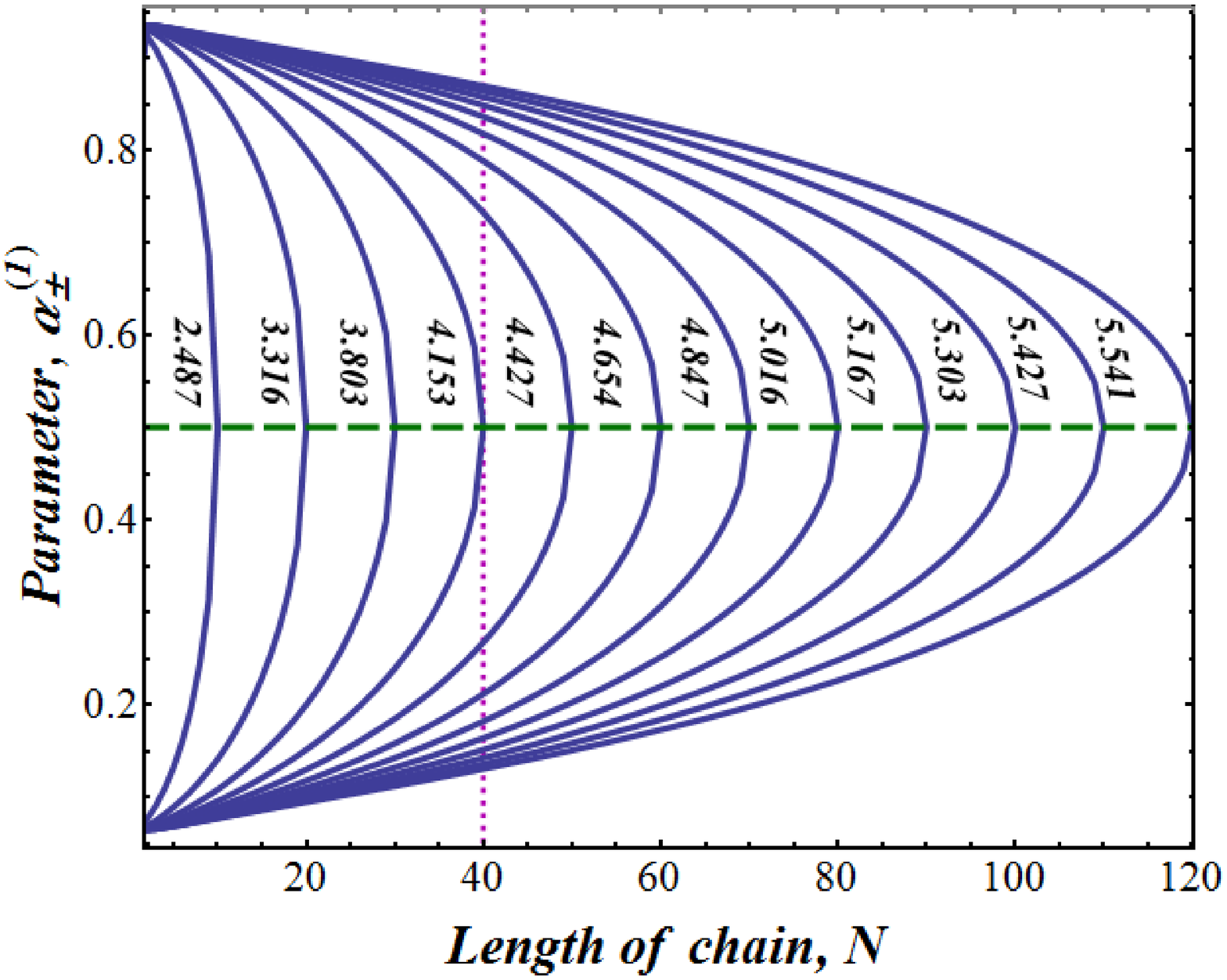,
  scale=0.33
   ,angle=0%270
}
\caption{
The creatable states  with the low coherence intensity.   
(a) The polarization $I^{(1)}_{-}$ (curves below  the dash-line $I^{(1)}_c$), 
$I^{(1)}_{+}$ (curves above  the dash-line $I^{(1)}_c$).
(b) The associated  values of the control parameter $\alpha$,
 $\alpha^{(1)}_{-}$ (curves below the dash-line $\alpha^{(1)}_c=\frac{1}{2}$), $\alpha^{(1)}_{+}$ 
 (curves above  the dash-line $\alpha^{(1)}_c$).  
All parameters are given as functions of the chain length 
$N$ for $J_{min}=0.01$.
Each curve corresponds to the particular $b$ from  set (\ref{bb}),
where  $n$ increases 
from the left to the right lines in this figure, for instance,  $b^{(1)}(10) =2.487$.}
  \label{Fig:q} 
\end{figure*}

For a given ${b}$, the intensity $J$ reaches its prescribed (registrable)  value $J_{min}$ for the  two values of $\alpha$:
\begin{eqnarray}\label{alp1pi}
\cos(\alpha^{(1)}_{\pm} \pi) =\mp \sqrt{1-\frac{4 J_{min}}{R^2} 
\left(\tanh\frac{{b}}{2}\right)^{2(1-N)}}.
\end{eqnarray}
Apparently, the expression under the square root must be non-negative, which holds for 
$
{b} \ge {b}^{(1)} 
$,
where the critical value $b^{(1)}$ corresponds to zero under the square root in expression (\ref{alp1pi}):
\begin{eqnarray}\label{bet1}
\tanh\frac{{b}^{(1)}}{2} = \left(\frac{4 J_{min}}{R^2}\right)^{\frac{1}{2(N-1)}}.
\end{eqnarray}
Next, substituting eq.(\ref{alp1pi}) into (\ref{IJ}), we eliminate dependence on $\alpha$ obtaining
%,
%eliminating $\cos(\alpha^{(1)} \pi)$ from system (\ref{IJ}) by means of eq.(\ref{alp1pi}) we obtain
\begin{eqnarray}\label{II2}
I^{(1)}_{\pm}=\frac{1}{2}\left(
(1-R^2)\tanh\frac{{b}}{2} \pm R \sqrt{R^2 - 4 J_{min}
\left(\tanh\frac{{b}}{2}\right)^{2(1-N)}}
\right)
\end{eqnarray}
The maximal $I^{(1)}_{+}$  and  minimal   $I^{(1)}_{-}$ values  correspond to, respectively,
the upper and lower signs in formula (\ref{II2}) 
(these parameters are related to, respectively, the right  and  left 
corners of the bell-shaped region).
Apparently, at $b=b^{(1)}$, the minimal  and  maximal 
values of coherence intensity coincide: $I^{(1)}_{-}=I^{(1)}_{+}= I^{(1)}_c$, where
\begin{eqnarray}\label{Imax22}
I^{(1)}_c=\frac{(1-R^2)}{2}\left(\frac{4 J_{min}}{R^2}\right)^{\frac{1}{2(N-1)}},
\end{eqnarray}
and the appropriate parameter $\alpha$ equals  $\alpha^{(1)}_c=\frac{1}{2}$, which follows from 
eq.(\ref{alp1pi}) at $b=b^{(1)}$. 
Parameters $I^{(1)}_{-}$, $I^{(1)}_{+}$  
together with the associated  values of the control parameters 
 $\alpha^{(1)}_{-}$, $\alpha^{(1)}_{+}$ as 
functions of the chain length 
$N$ for different values of $b$ in the list
\begin{eqnarray}\label{bb}
{b}=b^{(1)}(10 n), \;\;n=1,2,\dots,
\end{eqnarray}
and  $J_{min}=0.01$
are depicted in Fig.\ref{Fig:q}. 
The critical intensity $I^{(1)}_c$ and the critical parameter  $\alpha^{(1)}_c$ are  shown, respectively, in
Figs.\ref{Fig:q}a and \ref{Fig:q}b by the dash-lines. 
Parameters   ${b}^{(1)}$ and $I^{(1)}_c$
can be considered as the minimal required value of $b$ (or the maximal allowed  temperature) and the associated 
minimal value of the polarization required for creating the conventional value $J_{min}$ of 
the coherence intensity.

The interpretation  of Fig.\ref{Fig:q} is evident. The vertical line corresponding 
to the given chain length $N$ (see the vertical dot-lines in  Figs.\ref{Fig:q}a  and \ref{Fig:q}b  for $N=40$) 
crosses each line $b\ge  b^{(1)}(N)$ at the two points: $I^{(1)}_+(b)$ and 
$I^{(1)}_-(b)$ in Fig.\ref{Fig:q}a and $\alpha^{(1)}_+(b)$ and 
$\alpha^{(1)}_-(b)$ in Fig.\ref{Fig:q}b (these points are not marked in Fig.\ref{Fig:q}). These cross-points 
give the intervals of the
polarization and the appropriate intervals of the parameter $\alpha$ allowing us to create
the measurable  coherence intensity $J\ge J_{min}$.

  \subsection{Integral characteristics: fidelity of  remote state creation}
  \label{Section:IntChar}
\begin{figure*}
\epsfig{file=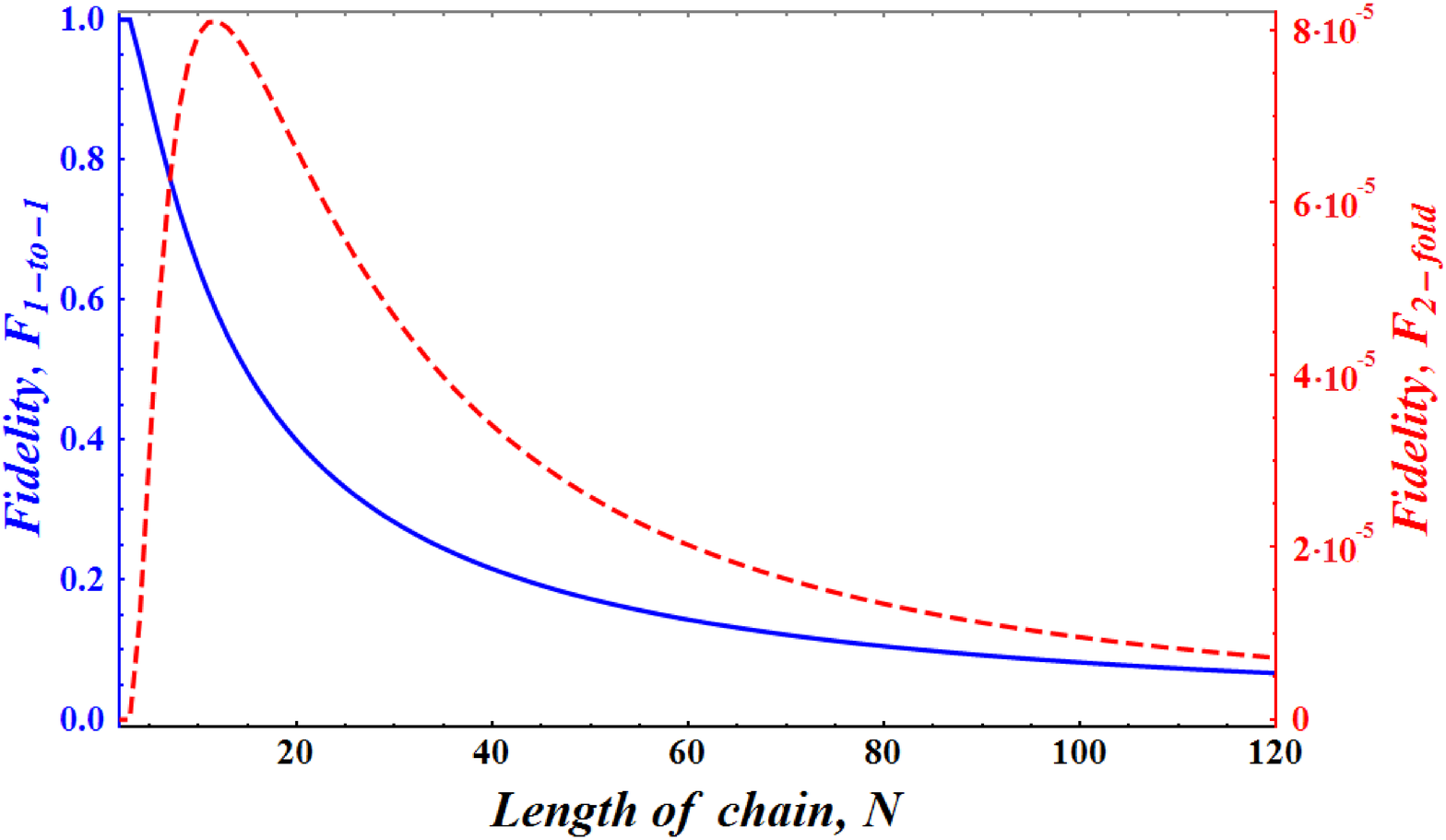,
  scale=0.3
   ,angle=0%270
   }
\caption{
The creation fidelities  of the one-to-one (bold line)  and the two-to-one (dash-line) 
mapped states are shown as functions of the chain length $N$. 
The fidelity of the  two-to-one mapped region is much less than that of the  
one-to-one mapped region and has the maximum at $N=12$. 
}
  \label{Fig:int} 
\end{figure*}

{  \subsubsection{Fidelity of remote state creation}}
  As  characteristics of the remote state creation, we propose the following integral characteristics, which 
  we refer to as the 
  fidelity of the remote state creation $F(N)$
  (by analogy with the similar characteristics of the state transfer \cite{Bose}) and define by 
  the following general formula:
  \begin{eqnarray}
  F(N)= \frac{S_{creatable}}{S_{receiver}}
  \end{eqnarray}
  where $S_{creatable}$ is the area of the creatable region and $S_{receiver}$ is 
  the area of the receiver's state space. 
  In turn,  $S_{creatable}$ can be splitt into two parts:
  the area of the one-to-one mapped creatable sub-region 
  $S_{1-to-1}$ and the area of the two-to-one mapped creatable sub-region 
  $S_{2-fold}$, so that 
  \begin{eqnarray}
  &&
  F(N)=F_{1-to-1}(N) + F_{2-fold}(N),\\\nonumber
  &&F_{1-to-1}(N)=\frac{S_{1-to-1}}{S_{receiver}},\;\;\; 
  F_{2-fold}(N)=\frac{S_{2-fold}}{S_{receiver}}.
  \end{eqnarray}
  In our case, $S_{receiver}$ can be calculated analytically using the boundary of region 
  (\ref{boundary}) $J=\frac{1}{4}-I^2$ and integrating it  over the interval
  $-\frac{1}{2}\le  I \le \frac{1}{2}$:
  \begin{eqnarray}
  S_{receiver} = \frac{1}{6}.
  \end{eqnarray}
  The function 
$S_{1-to-1}$ can be also calculated analytically eliminating $\alpha$  from 
eq.(\ref{liminfJ}) by means of eq.(\ref{liminfI}) 
and integrating the obtained coherence intensity $J_\infty$  as a  function of polarization $I_\infty$,
\begin{eqnarray}\label{JS}
J_\infty=\frac{1}{4 R^2} (1-2 I_\infty)(2 I_\infty + 2 R^2-1),
\end{eqnarray}
over the interval $I_c \le  I_\infty \le \frac{1}{2}$:
  \begin{eqnarray}
  S_{1-to-1} = \frac{R^4}{6}.
  \end{eqnarray}
Finally, the function $S_{2-fold}(N)$ can be calculated numerically using the formulas (\ref{Iextr}) 
and (\ref{Jextr}) for the upper boundary of this region 
(over  interval (\ref{intF2fold1})) and (\ref{JS}) for the right boundary
(over  interval (\ref{intF2fold2})):
\begin{eqnarray}\label{S2fold}
S_{2-fold}=\int\limits_{-\frac{R^2}{2}}^{I^{br}_\infty} d I \; J^{br}_+(I) -\frac{1}{4 R^2}
\int\limits_{I_c}^{I^{br}_\infty} d I \;(1-2 I)(2 I + 2 R^2-1),
\end{eqnarray}
where $I_c$ and  $I^{br}_\infty$ are
defined in eqs.(\ref{Ic}) and (\ref{Iextr2}) respectively. Expression for  $J^{br}_+(I)$ in 
integral (\ref{S2fold}) follows   from eq.(\ref{Jextr}) after eliminating 
$\tanh\frac{b^{br}}{2}$ by means of eq.(\ref{Iextr}).
The fidelities  $F_{1-to-1}$ and $F_{2-fold}$ as functions of $N$ are depicted in Fig.\ref{Fig:int}.
This figure shows the smallness of   the two-to-one mapped sub-region  in comparison 
with the one-to-one mapped one in our model. We can also see the  maximum of the fidelity $F_{2-fold}(N)$  at $N=12$.

{ \subsubsection{Average creatable polarization and coherence intensity as functions of temperature}
The interesting characteristics are the polarization and the coherence intensity averaged over 
the initial pure state of the sender:
\begin{eqnarray}\label{Iav}
\bar I(b) &=& \int\limits_{0}^1 I(\alpha,b) d\alpha = \frac{1-R^2}{2} \tanh\frac{b}{2},
\\\label{Jav}
\bar J(b) &=& \int\limits_{0}^1 J(\alpha,b) d\alpha =\frac{R^2}{8} \left(\tanh\frac{b}{2}\right)^{2(N-1)} .
\end{eqnarray}
where $I$ and $J$ are defined in eqs.(\ref{IJ}) and (\ref{Jdef}), respectively. 
Equations (\ref{Iav}) and (\ref{Jav}) show that  both the mean polarization and the mean coherence intensity 
increase with $b$. Therewith, the mean  polarization increases  also with $N$, while the mean 
coherence intensity decreases with $N$.
The families of curves 
$\bar I(b)$ and $\bar J(b)$ are represented in Fig.\ref{Fig:avr} for different $N$.
\begin{figure*}
\epsfig{file=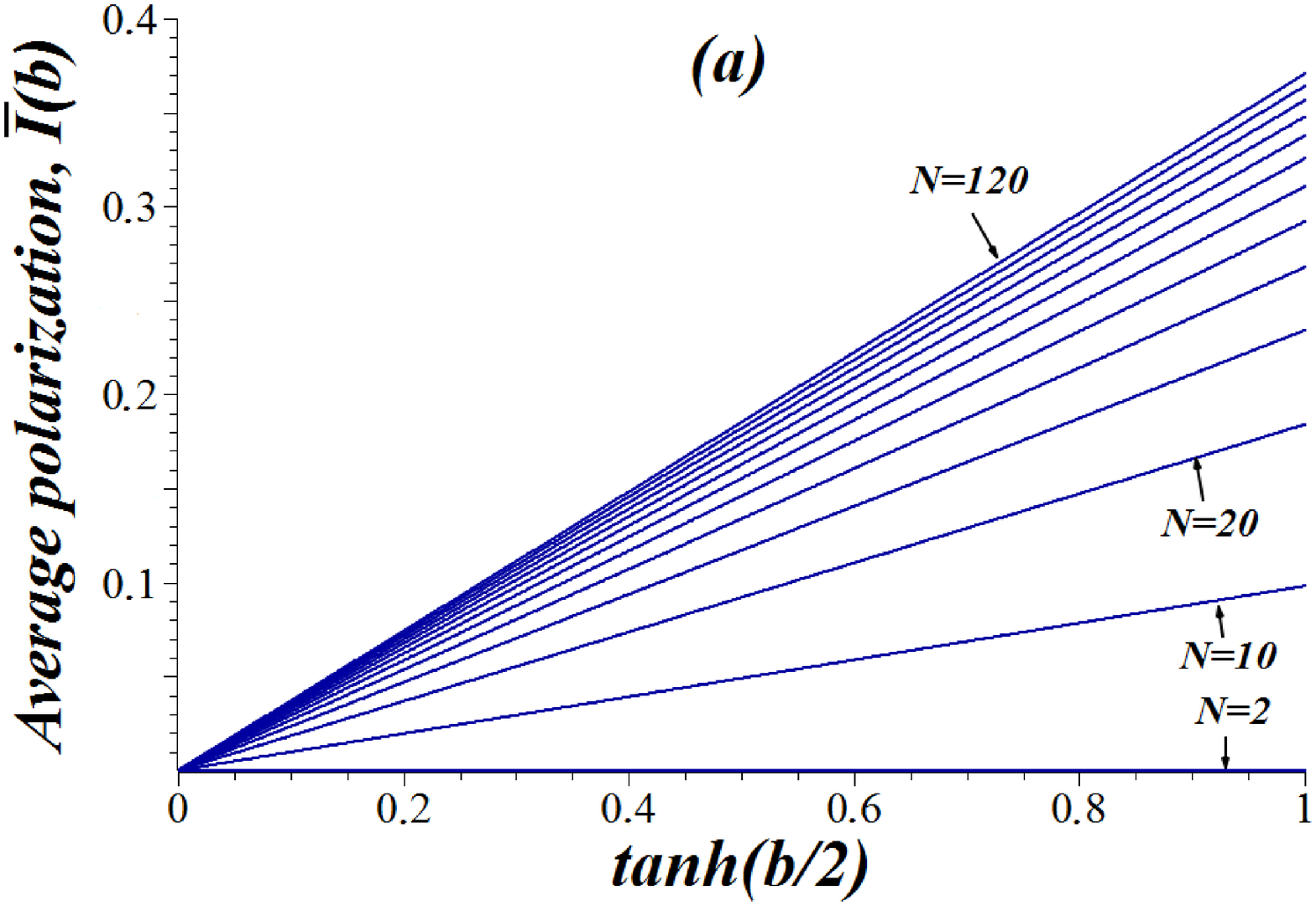,
  scale=0.3
   ,angle=0%270
   }
  \epsfig{file=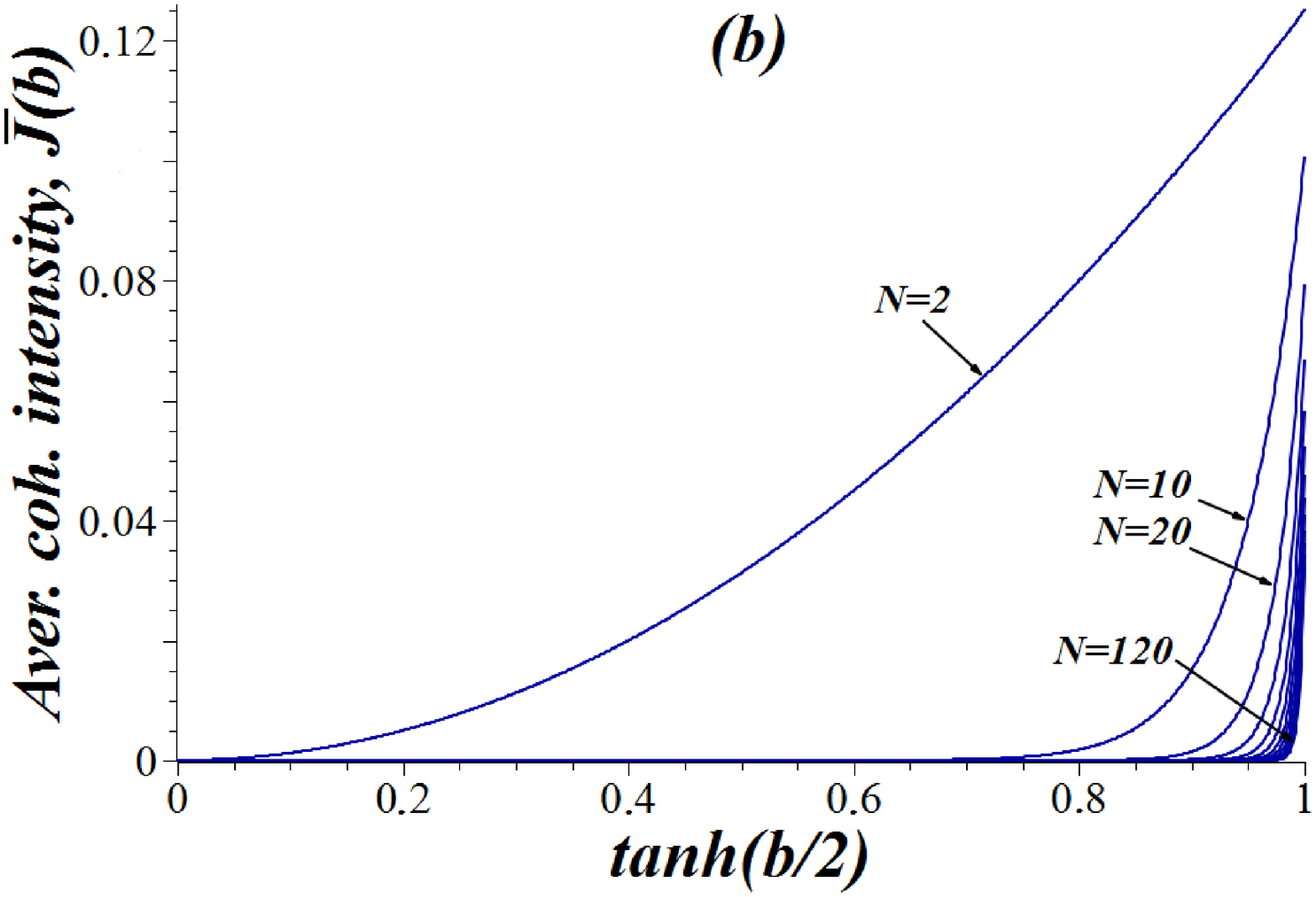,
  scale=0.3
   ,angle=0%270
   }
\caption{{The polarization (a) and the
coherence intensities (b) averaged over the control parameter $\alpha$ are shown 
as functions of the inverse temperature $b$ for chains of different lengths 
$N$, $N=2,10,20,\dots,120$.} 
}
  \label{Fig:avr} 
\end{figure*}

}

%%%%%%%%%%%%%%%%%%%%%%%%%%%%%%
\section{Parametrization of  receiver's density matrix in terms of independent 
eigenvalue-eigenvector parameters}
\label{Section:param1}
Although the physically motivated parametrization considered in Sec.\ref{Section:coh} allows us to give 
a  comprehensive description of the creatable region together with its  physical interpretation 
in terms of polarization and  coherence intensity, we consider another 
parametrization 
in terms of the 
receiver's eigenvalues and eigenvectors. 
The motivation for this parametrization is the comparison of our model 
of remote state  creation with that described in \cite{BZ_2015}, 
where the homogeneous chain with  the pure initial state of the two-qubit 
sender and  ground initial state of the rest system was considered, therewith the receiver was a one-qubit subsystem 
as well.

Thus, we represent state  (\ref{rhoN0})  of the receiver 
in the following factorized form:
\begin{eqnarray}\label{ULU}
\rho^N =  U^N \Lambda^B (U^N)^+,
\end{eqnarray}
where 
$\Lambda^N$ is the diagonal matrix of  eigenvalues and $U^N$ is the matrix of  eigenvectors,
which read as follows in our case: 
\begin{eqnarray}
\label{Lambda}
&&
\Lambda^N={\mbox{diag}}(\lambda,1-\lambda),\\\label{U}
&&
U^N=
\left(\begin{array}{cc}
\cos \frac{\beta_1 \pi}{2} & -e^{-2 i \beta_2\pi} \sin \frac{\beta_1\pi}{2} \cr
e^{2 i \beta_2\pi} \sin \frac{\beta_1\pi}{2}  & \cos \frac{\beta_1\pi}{2}.
\end{array}\right)
\end{eqnarray}
with $\lambda$ and $\beta_i$ ($i=1,2$) varying  inside of the intervals 
\begin{eqnarray}
\label{lamint}
&&
\frac{1}{2} \le \lambda \le 1, \\\label{betint}
&&
0\le \beta_i\le 1,\;\;i=1,2.
\end{eqnarray}
Comparing eq.(\ref{ULU}) with eq.(\ref{rhoN0}) we establish the relation between the control parameters 
$(\alpha,{b},\phi)$ and the creatable parameters $(\lambda,\beta_1,\beta_2)$:
\begin{eqnarray}\label{lam}
&&
\lambda= \frac{1}{2} \left( 1+ \sqrt{(2 \rho^N_{11} -1)^2 + 4 |\rho_{12}|^2}\right)=\\\nonumber
&&
\frac{1}{2}\left(
1+\sqrt{1+\Delta_0}\Big|R^2\cos(\alpha\pi) -(1-R^2)\tanh\frac{{b}}{2} \Big|\right),
%&&
%\frac{1}{2}\left(
%1 + \sqrt{\Big(R^2 \cos (\alpha \pi) - \tanh\frac{{b}}{2} (1-R^2)\Big)^2+  R^2 \sin^2 (\alpha \pi)\Big(\tanh\frac{{b}}{2}\Big)^{2(N-1)}}
%\right)
\\\label{bet}
&&
\cos(\beta_1\pi) = \frac{2\rho^N_{11}-1}{2\lambda -1} =\\\nonumber
&&
-\frac{1}{\sqrt{1+\Delta_0}}{\mbox{sign}}(R^2\cos(\alpha\pi) -(1-R^2) \tanh\frac{{b}}{2} )
%&&
%-\frac{R^2 \cos (\alpha \pi) - \tanh\frac{{b}}{2} (1-R^2)}{\sqrt{(R^2 \cos (\alpha \pi) - 
%\tanh\frac{{b}}{2} (1-R^2))^2  + 
%R^2 \sin^2 (\alpha \pi)\Big(\tanh\frac{{b}}{2}\Big)^{2(N-1)}}}
\\\label{bet2}
&&
\beta_2=\Phi_N(\tau)+\phi +  \frac{1}{2} (N-1),
\end{eqnarray}
where
\begin{eqnarray}\label{Delta0}
\Delta_{0}= \frac{R^2\sin^2(\alpha\pi) \Big(\tanh\frac{{b}}{2}\Big)^{2(N-1)}}{
(R^2 \cos (\alpha \pi) - (1-R^2)\tanh\frac{{b}}{2} )^2} 
\end{eqnarray}
Intervals (\ref{lamint}) and (\ref{betint}) cover the whole state-space of the receiver.
We see that any parameter $\beta_2 $ can be created at the fixed time instant $\tau$ using the control parameter 
 $\phi$: $\phi =\beta_2(\tau)-\Phi_N(\tau)-  \frac{ (N-1)}{2} $. For this reason, similar to Sec.\ref{Section:coh},
 we consider  density matrix (\ref{rhoN}) instead of (\ref{rhoN0}) and thus, 
studying  the map (control parameters) $\to$ (creatable parameters),
we turn to the reduced  map
\begin{eqnarray}\label{ablambet}
(\alpha,b)\stackrel{(\ref{lam},\ref{bet})}{\to}(\lambda,\beta_1)
\end{eqnarray}
instead of the complete map 
$(\alpha,b,\phi)\stackrel{(\ref{lam},\ref{bet},\ref{bet2})}{\to}(\lambda,\beta_1,\beta_2)$.

Notice that 
two independent  
pairs of the control parameters $\{I,J\}$
and $\{\lambda,\beta_1\}$   are  related by the  following  one-to-one map:
\begin{eqnarray}\label{IJlambet}
\begin{array}{l}
\lambda = \frac{1}{2} +\sqrt{I^2 + J}\cr
\cos\beta_1\pi =\frac{I}{\sqrt{I^2 + J}}
\end{array} \;\;\Leftrightarrow \;\;
\begin{array}{l}
I=\Big(\lambda-\frac{1}{2}\Big) \cos(\beta_1 \pi)\cr
\sqrt{J}=\Big(\lambda-\frac{1}{2}\Big) \sin(\beta_1 \pi)
\end{array}.
\end{eqnarray}
Formulas in (\ref{IJlambet}) suggest us to combine all these parameters in single relation  
introducing the following complex  function $\Xi$:
\begin{eqnarray}
\Xi \equiv I + i \sqrt{J} = \Big(\lambda-\frac{1}{2}\Big) e^{i\beta_1\pi}.
\end{eqnarray}

\subsection{Creatable regions for  homogeneous chains of different lengths}

\begin{figure*}
   \epsfig{file=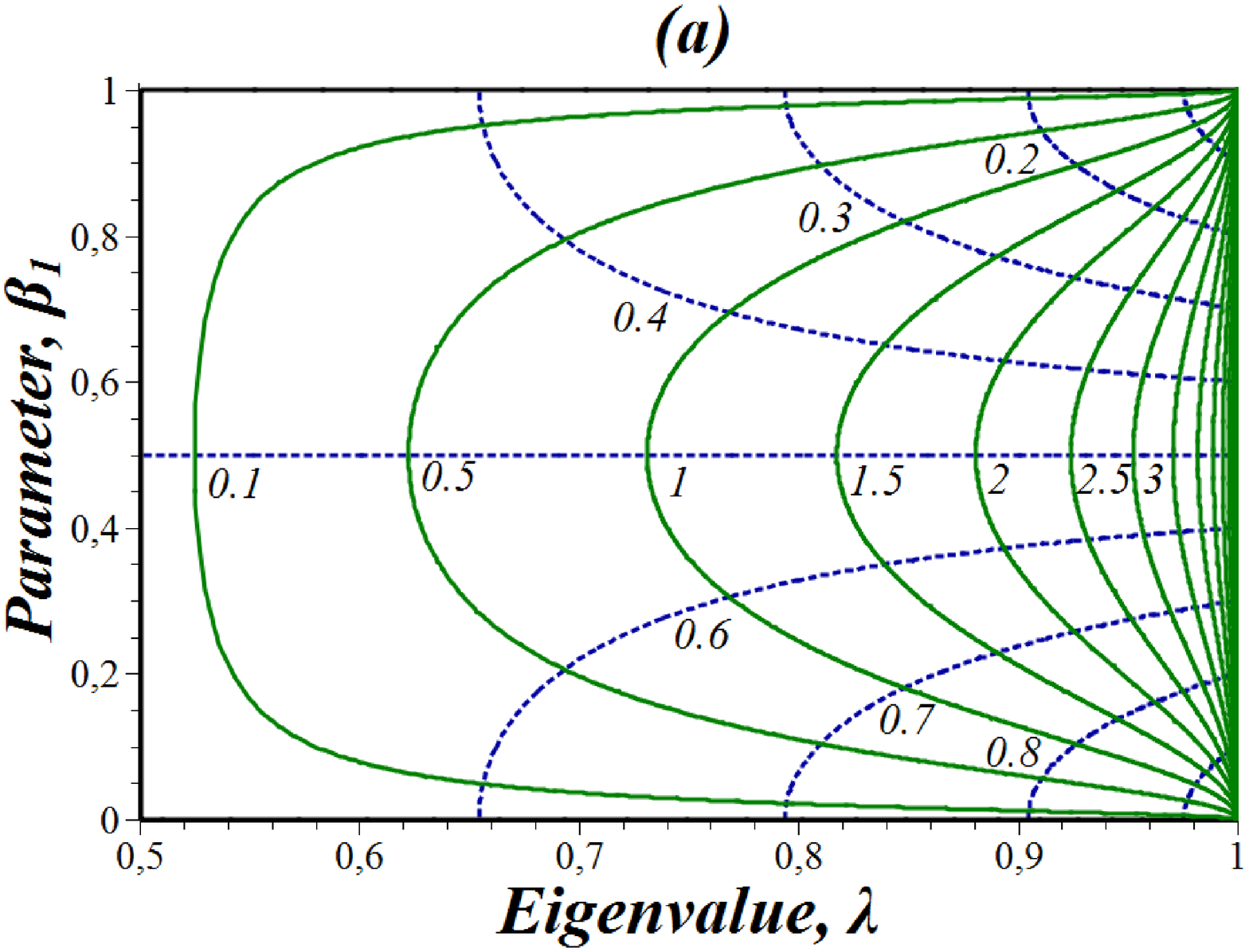,
  scale=0.3
   ,angle=0%270
}
\epsfig{file=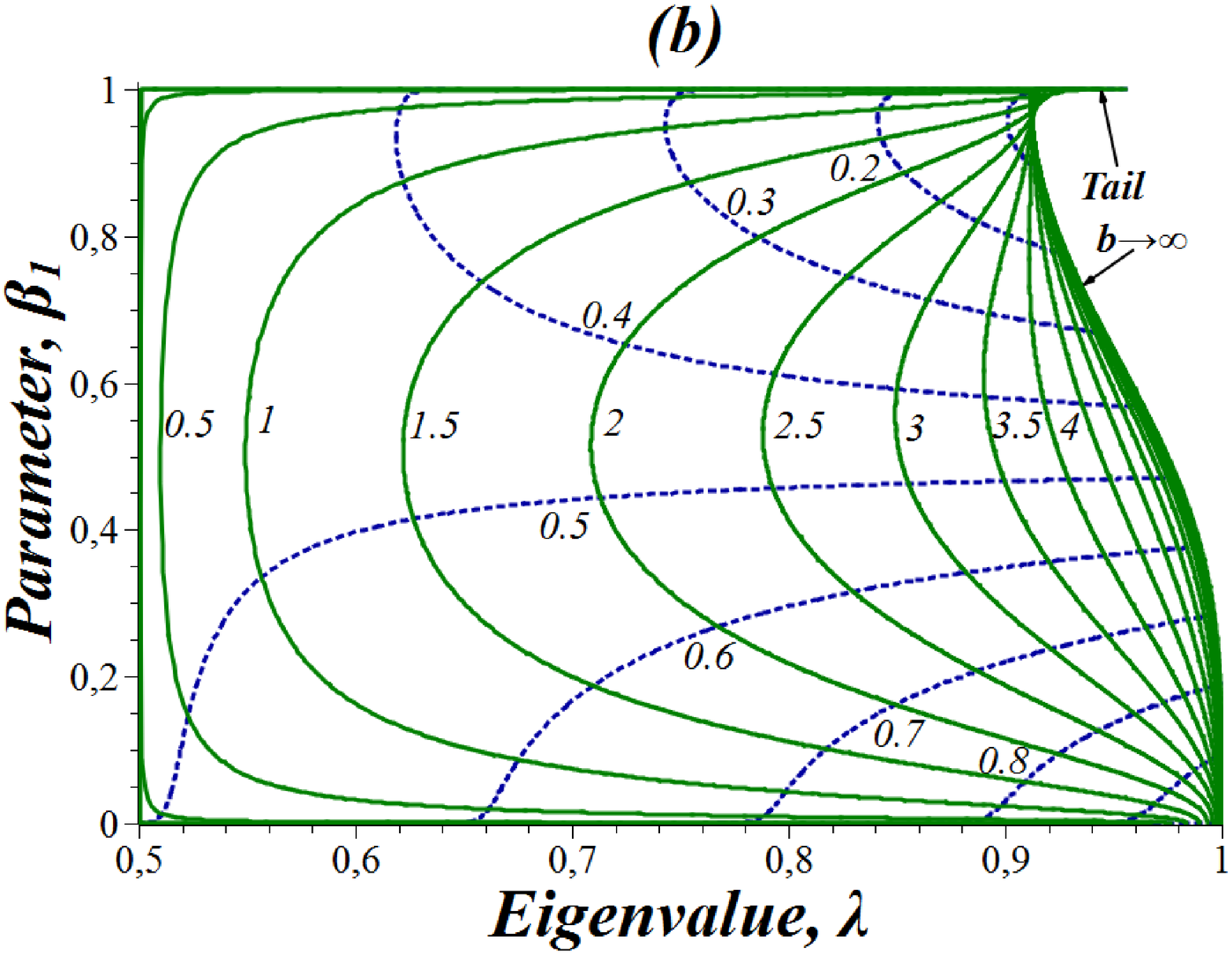,
  scale=0.3
   ,angle=0%270
}\newline

\epsfig{file=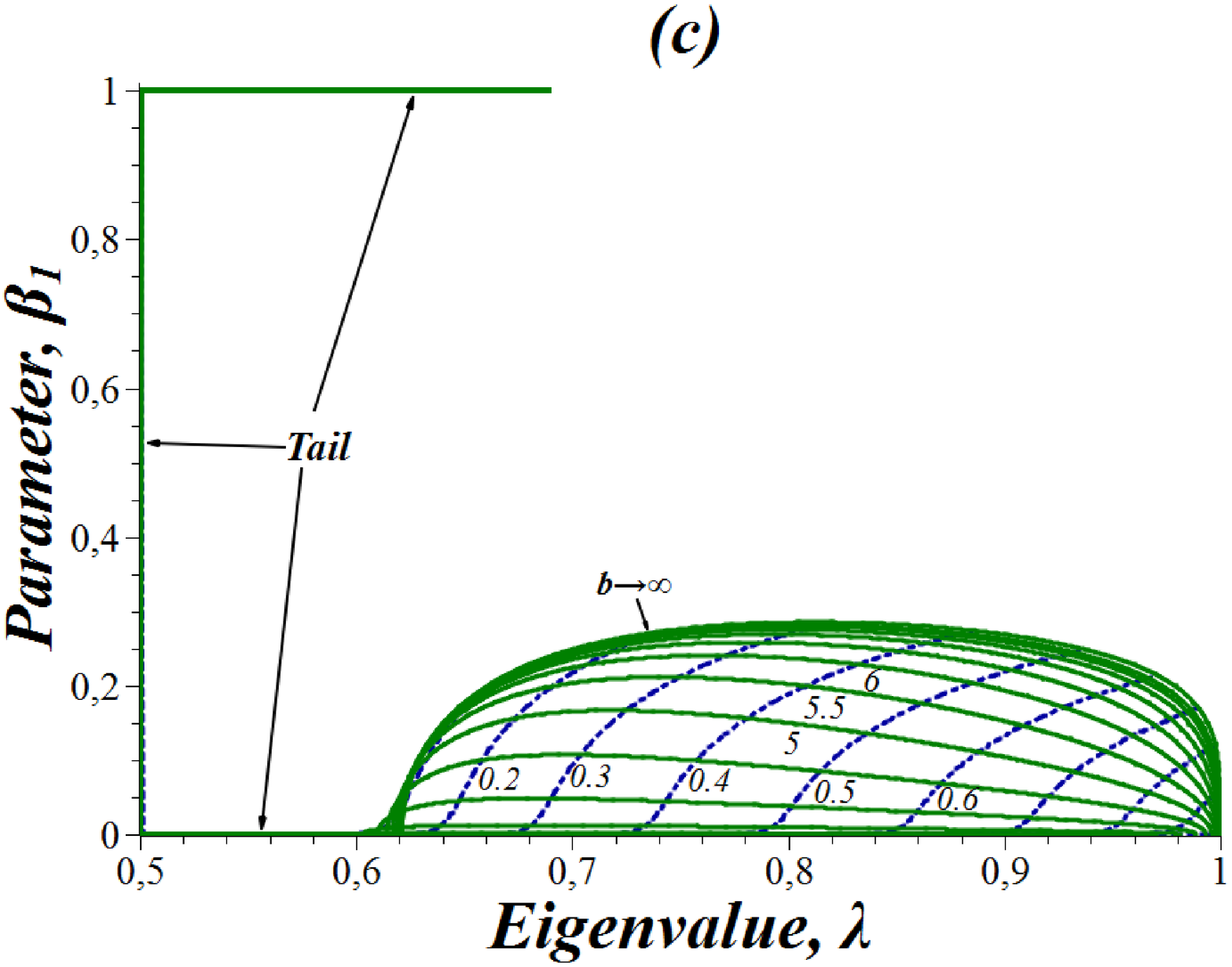,
  scale=0.3
   ,angle=0%270
}
\epsfig{file=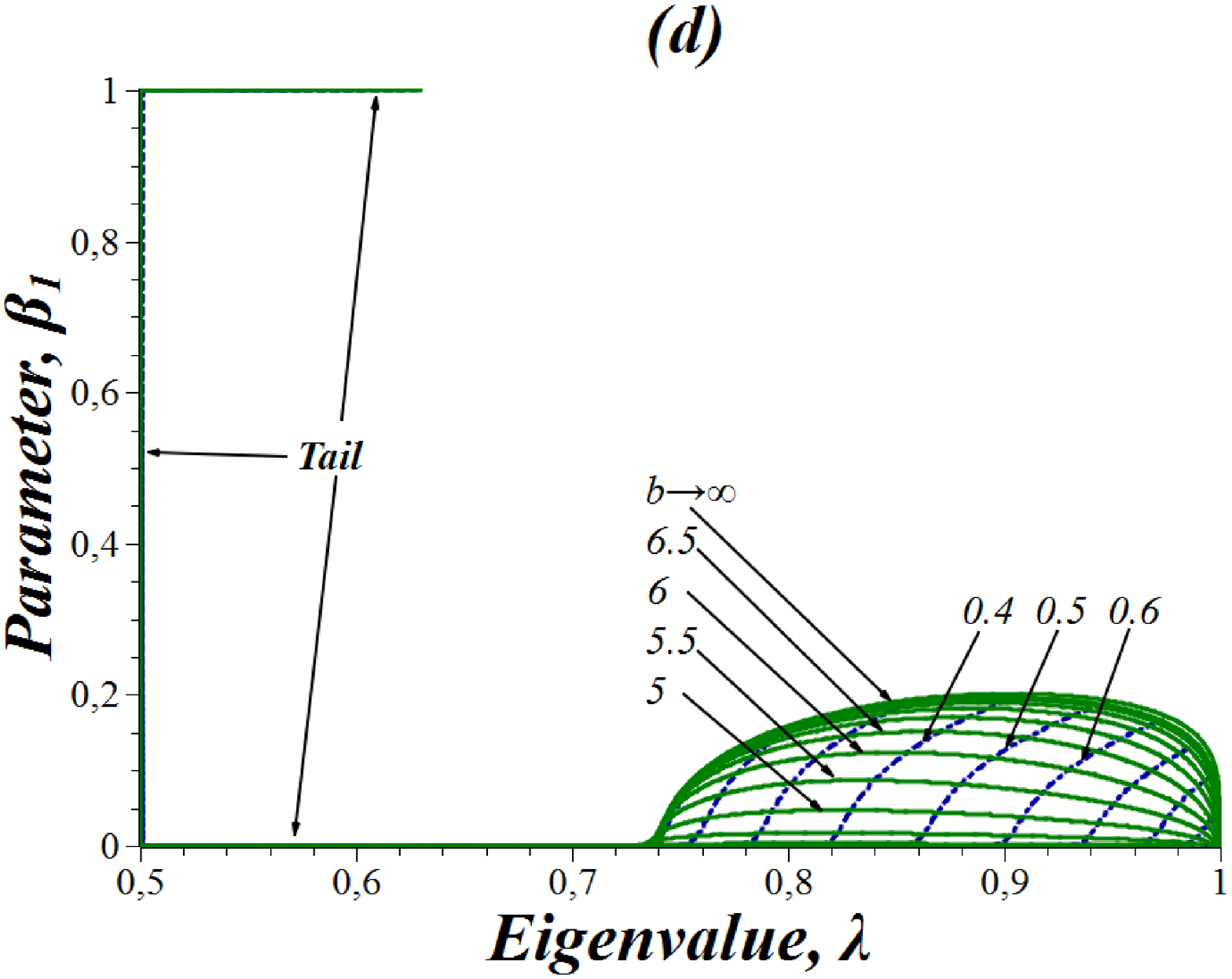,
  scale=0.3
   ,angle=0%270
   }
\caption{The creatable regions in the space of the creatable parameters $\lambda$ and $\beta_1$ for
(a) $N=2$, (b) $N=6$, (c) $N=60$, and (d) $N=120$. General behavior of the 
curves $b=const$ (solid lines) and $\alpha=const$ (dash-lines)
is similar to that of curves in Fig.\ref{Fig:regPolN2to120}. In particular, the curves 
$b=const$ concentrate near the curve $b\to\infty$. With an increase in $N$, their 
density near $b=0$ increases as well. For $N=2$, the line $b\to\infty$ coincides  with the right
coordinate line of the parameter $\beta_1$.
The tail of the states with vanishing $I$ 
is well formed in the cases $N=6$, 60 and 120 (shown in figs. (b)-(d)).
This tail corresponds to  $\alpha \to 0$ (any $b$),
similar to  Fig.\ref{Fig:regPolN2to120}.
The unavailable region appears if $N>3$.}
  \label{Fig:regN2to120} 
\end{figure*}

Here we describe the creatable regions in chains of different lengths. 
The maximal region corresponds to 
$N=2$ or $3$, as  shown in Fig.\ref{Fig:regN2to120}a.
The common feature of the creatable regions in Fig.\ref{Fig:regN2to120}  
is the tail of states with $\beta_1 \to 0,1$ appearing in chains of the length $N>3$, 
similar to the case of physical parametrization shown in 
Fig.\ref{Fig:regPolN2to120}. 
But now the tail is ''crooked'' for large $N$, as shown in 
Fig.\ref{Fig:regN2to120}c,d. 

Formulas (\ref{IJlambet}) allow us to rewrite all characteristics of the creatable region found in 
Sec.\ref{Section:coh} in terms of the parameters $\lambda$ and $\beta_1$. 
In particular, 
using  formulas (\ref{IJlambet}), we conclude that 
\begin{eqnarray}
\beta_1\to 0  \;\;\Leftrightarrow \;\; I >0,\; J\to 0,\\\nonumber
\beta_1\to 1  \;\;\Leftrightarrow \;\; I <0,\; J\to 0.\\\nonumber
\end{eqnarray}
Therewith  there is a particular case $I=0$  yielding
\begin{eqnarray}
\beta_1^{cr} = \frac{1}{2}  \;\;\Leftrightarrow \;\;  I =0.
\end{eqnarray}
Now, eq.(\ref{bet}) 
with $\beta_1=\frac{1}{2}$  yields  eq.(\ref{cr1}),
while eq.(\ref{lam}) results in  
\begin{eqnarray}
&&
\lambda^{(0)}=\frac{1}{2}\left(
1+\sqrt{R^4-(1-R^2)^2\tanh^2\frac{{b^{(0)}}}{2}} \Big(\tanh\frac{{b^{(0)}}}{2}\Big)^{N-1}
\right).
\end{eqnarray}
Eq.(\ref{cr1}) holds for all $b$ if condition (\ref{conRR}) is satisfied, which leads to 
the critical value $N_c= 34$ 
 obtained in   Sec.\ref{Section:cohNc} and in Ref.\cite{BZ_2015}. 
Thus, inequality (\ref{constrIJ}) holds for all $b$ if $N \le N^c$.
For  $N>N^c$,    condition (\ref{constrIJ}) is satisfied if 
\begin{eqnarray}
b \le {b^{(0)}}=2\; {\mbox{atanh}}\frac{R^2}{1-R^2}.
\end{eqnarray}
 This means that the curves ${b}=const>{b^{(0)}}$ 
on the graphs 
do not reach the upper boundary $\beta_1=1$, while the other curves (${b}=const\le {b^{(0)}}$) 
reach it.
Notice also that the state with $\lambda=\frac{1}{2}$ assumes arbitrary $\beta_1$. Thus,
the whole vertical line $\lambda=\frac{1}{2}$ in
Figs.\ref{Fig:regN2to120} represents the same state.

Similarly to Fig.\ref{Fig:regPolN2to120},
the whole creatable region is divided into two parts: the image of  one-to-one map (\ref{ablambet},\ref{OneToOne}) 
(the main part of the creatable region)
and the image of  
two-to-one map (\ref{ablambet},\ref{TwoFold}), which is the tail together  with the 
close neighborhood of the corner where
this tail attaches to the 
main part of the creatable region.  
These two subregions are separated one from another by   the curve ${b}\to\infty$ (see
Fig.\ref{Fig:regN2to120}):
\begin{eqnarray}
&&\lambda|_{{b}\to\infty} =\frac{1}{2}\left(
1 + \sqrt{(R^2(\cos(\alpha\pi) +1) - 1)^2 + R^2 \sin^2(\alpha \pi) }
\right),\\
&&\cos(\beta_1 \pi|)|_{{b}\to\infty}=
-\frac{R^2(\cos(\alpha\pi) +1) - 1}{\sqrt{(R^2(\cos(\alpha\pi) +1) - 1)^2 + R^2 \sin^2(\alpha \pi) }}.
\end{eqnarray}

With an increase in $N$, the creatable region vanishes. Here we shall emphasize the important difference between 
the model of Ref.\cite{BZ_2015} and our one. 
Unlike that model, any eigenvalue can be simply created in our case as 
the parameter associated with the classical limit.
On the contrary, 
the parameter $\beta_1$, characterizing the eigenvector, 
has restrictions for its creatable values in both models.

\section{Conclusion}
\label{Section:conclusions}

We consider the  remote creation of  the polarization and the intensity of the first-order coherence 
in a spin-1/2 chain with the physically motivated initial condition: the pure state of 
the one-qubit sender and the thermodynamic equilibrium state 
of the rest nodes. Therewith, the
polarization is basically responsible for the classical effects (the diagonal 
part of the  density matrix) while the
coherence intensity is responsible for the quantum effects (the non-diagonal part of the density matrix).
Using this physical parametrization, we obtain the following properties and characteristics of 
the creatable region of 
the receiver state-space in our model.
\begin{enumerate}
\item
At the fixed temperature (parameter $b$), 
we can use the parameter $\alpha$  of the sender's initial state as the control parameter 
 varying the values of the polarization and the coherence intensity (i.e, the parameter $\alpha$ moves the state $(I,J)$
 along the chosen solid curve 
 in Fig.\ref{Fig:regPolN2to120})). Thus, the temperature is not the only parameter 
 responsible for the remotely creatable polarization and coherence intensity (Sec.\ref{Section:CRREG}).
\item
The creatable region is divided into two subregions covered by, respectively,  one-to-one map 
(\ref{abIJ},\ref{OneToOne}) and two-to-one map (\ref{abIJ},\ref{TwoFold}), therewith the biggest subregion is 
created by the one-to-one map, which is, essentially, the bell-shaped regions in Fig.\ref{Fig:regPolN2to120}. 
The boundaries of these subregions are described, see 
Secs.\ref{Section:onetoone} and \ref{Section:twofold}. 
\item
Any state $(I,J)$ from the two-to-one mapped  subregion (except  the upper boundary ($I^{br}_+,J^{br}_+$) 
given in (\ref{Iextr},\ref{Jextr})) can be created using the two different pairs of 
the control parameters $(\alpha,b)$. In particular, this can be achieved fixing $\alpha$ 
(i.e., the pure state of 
the one-node sender) and varying  the temperature or vise-versa (Sec.\ref{Section:twofold}). 
\item
The tail of states with vanishing coherence intensity and large  polarization 
is well formed 
to the left  of the bell-shaped region. This tail is referred to the two-to-one mapped subregion, 
see Sec.\ref{Section:tail}.
\item
The states with zero polarization $I$ and large coherence intensity $J$ are described in Sec.\ref{Section:cohNc}. 
Such  states are creatable only in short chains. With an increase in $N$, the significant value of the 
coherence intensity can be created 
only together with the large polarization. 
\item
On the contrary, the large  polarization can be created without the coherence intensity in the long chains, 
which justifies the classical origin of the polarization, see Sec.\ref{Section:largepol}. Moreover, as $N\to\infty$,
the creatable intensity shrinks to the interval $0\le I\le \frac{1}{2}$.
\item
For a given $N$ and $b>b^{(1)}$, the measurable value of the  coherence intensity $J\ge J_{min}$ can be achieved only for 
$\alpha$  chosen  inside  of the interval $\alpha^{(1)}_- \le \alpha\le \alpha^{(1)}_+$ 
(and consequently, the value of creatable  polarization $I$ is  inside of the 
interval $I^{(1)}_- \le I\le I^{(1)}_+$), Sec.\ref{Section:largepol}.
\item
We introduce the integral characteristics of the remotely created region (as well as  of the both
 one-to-one and two-to-one mapped subregions)
as the ratio of the area of the creatable (sub)region to the area of the whole receiver state-space,
see Sec.\ref{Section:IntChar}. By analogy with the pure state transfer, we refer to this characteristics as the
state-creation fidelity. The polarization and coherence intensity averaged over the 
sender's initial state are also found as functions of the inverse temperature $b$ and  chain length $N$.

\item
To simplify the comparison of our model with the state creation model proposed in \cite{BZ_2015}, 
we also consider the alternative parametrization in terms of the eigenvalue and 
eigenvectors of the receiver state in Sec.\ref{Section:param1}.
A principal distinctions of our model is the possibility to transfer 
any eigenvalue over the  homogeneous spin chain 
of any length $N$, while this length is restricted by the critical value  $N_c =34$
in \cite{BZ_2015}. 
\end{enumerate}

Thus we represent the analysis of the creatable polarization and coherence intensity, which are uniquely related with the state of 
one-qubit receiver. In turn, the state of multi-qubit receiver can be (partially)  characterized by the
intensities of multiple quantum coherences, 
for which the detection methods are well developed \cite{BMGP,F}. 

Now we underline several  problems of interest  prompted by our results.

First of all, the study of  the possibility to use a local tool on the receiver
site with the purpose to  increase the creatable space of the receiver (i.e., fidelity). 
Since any eigenvalue can be transfered, then, in principle, this 
problem can be solved  using the local unitary transformation on the receiver side.  However, 
the local transformations independent on the control parameters (the parameters $\alpha$ and $b$ 
in our case) would be  of general interest. The way  realizing  such the local transformations is not evident.

Next,  we must keep in mind the problem of remote control of 
quantum operations based on mixed states. In this regards, the two-to-one mapped subregion 
might be useful. In particular, an ambiguity of the 
control parameters creating the given state would be of possible interest in quantum cryptography 
\cite{GRTZ}, which is another applicability direction of quantum information devises
\cite{HBB,SCCS}.

In addition, one should remember that the parameter $b$ (the inverse temperature) is 
one of the control parameters, which is not a sender's local parameter, but the 
characteristics of the chain as a whole. Thus, we must be able to provide the constant temperature 
in the whole sample containing our spin chain. 

Notice also that the temperature is a 
global parameter of the whole chain.  This means that the remote state creation in our model partially  
 loses its control exclusively by the local parameters of the sender. However, since the  chain's established
 temperature is 
a known parameter on the receiver's site, we do not need to transfer this parameter through 
any classical communication channel.

This  work is partially  supported by the Russian Foundation for Basic Research, grants No.15-07-07928 and 
No.16-03-00056, and by the Program of RAS ''Element base of quantum computers'' (grant No.0089-2015-0220).

\section{Appendix A: Amplitude   $R_N(\tau)$ as a characteristics of  transmission line}

The nearest neighbor Hamiltonian (\ref{XY}) can be diagonalized using 
the Jordan-Wigner transformation method \cite{JW,CG}:
\begin{eqnarray}\label{Hdiag}
H=\sum_{k} \varepsilon_k \beta_k^+\beta_k ,\;\;
\varepsilon_k =  \cos(k) ,\;\;\displaystyle k=\frac{\pi n}{N+1}, \;\;\;n=1,2,\dots,N,
\end{eqnarray}
where   $\beta_j$ are the fermion operators, introduced in terms of  the   other fermion operators $c_j$  
using the Fourier transformation
\begin{eqnarray}\label{betc}
\beta_k = \sum_{j=1}^N g_k(j) c_j,
\end{eqnarray}
where the fermion operators  $c_j$ read
\begin{eqnarray}
c_j=(-2)^{j-1} I_{1z}I_{2z}\dots I_{(j-1)z} I^-_j.
\end{eqnarray}
Here
\begin{eqnarray}\label{g}
g_k(j)=\left(
\frac{2}{N+1}
\right)^{1/2} \sin(k j).
\end{eqnarray}
%with $\displaystyle k=\frac{\pi n}{N+1}$, $n=1,2,\dots,N$.
The  projection operators $I_{jz}$ can be represented as 
\begin{eqnarray}\label{Izc}
I_{jz} = c^+_j c_j -\frac{1}{2},\;\;\forall \; j.
\end{eqnarray}

Before  proceed  to the derivation of the density matrix evolution, we rewrite 
initial density matrix (\ref{initial}) in the following operator form
\begin{eqnarray}
\label{opform}
&&
\rho_0=\frac{1}{Z}(\frac{1}{2} E + (|a_0|^2-|a_1|^2) I_{z1} + a_0 a_1^* I^+_1 + 
a_1 a_0^* I^-_1) e^{-{b} I_{z1}} e^{{b} I_z}=\\\nonumber
&&
\frac{1}{Z}(A_1 E + A_2I_{z1} + A_3 I^+_1 + 
A_4 I^-_1)  e^{{b} I_z}
,
\end{eqnarray}
where $E$ is the $2\times 2$ unit operator,
\begin{eqnarray}
&&
Z=\left(2 \cosh\frac{{b}}{2}\right)^{N-1},\;\;I_z =\sum_{i=1}^N I_{zi},\\\nonumber
&&
A_1=\frac{1}{2}e^{-\frac{{b}}{2}} +|a_1|^2 \sinh\frac{{b}}{2},\;\;\; 
A_2=e^{-\frac{{b}}{2}} -2 |a_1|^2 \cosh\frac{{b}}{2},\\\nonumber
&&
A_3=a_0 a_1^* e^{\frac{{b}}{2}},\;\;\;A_4=a_0^* a_1 e^{-\frac{{b}}{2}}
.
\end{eqnarray}
Since $[H,I_z]=0$,
the evolution of the density matrix can be written as 
\begin{eqnarray}
\rho(\tau)=\frac{1}{Z}\sum_{i=1}^4 {r_i(\tau)} e^{{b} I_z},
\end{eqnarray}
with 
\begin{eqnarray}
&&
r_1(\tau)=A_1 ,\;\;
\\\nonumber
&&
r_2(\tau) = A_2\left(
-\frac{1}{2} + \sum_{k,k'=1}^N e^{-i \tau(\varepsilon_k-\varepsilon_{k'})} g_{1k}g_{1k'}\beta^+_k\beta_{k'}
\right),\\\nonumber
&&
r_3(\tau)=A_3\sum_{k=1}^N e^{-i \tau \varepsilon_k} g_{1k}  \beta^+_k,\;\;\;
r_4(\tau)=A_4\sum_{k=1}^N e^{i \tau \varepsilon_k} g_{1k}  \beta_k.
\end{eqnarray}
Reducing this matrix with respect to all the nodes except for the $N$th one and writing it in the basis 
$|0\rangle$, $|N\rangle$, 
we obtain the state of the last node:
\begin{eqnarray}\label{ArhoN}
\rho_N(\tau)= 
\left(\begin{array}{cc}\displaystyle
\frac{e^{\frac{{b}}{2}}}{2 \cosh\frac{{b}}{2}} + \frac{1}{2}\left(
\frac{e^{-\frac{{b}}{2}}}{\cosh\frac{{b}}{2}}- 2|a_1|^2\right) |f_N(\tau)|^2 & \displaystyle
(-\tanh\frac{{b}}{2})^{N-1} a_0 a_1^*
f^*_N(\tau)\cr\displaystyle
(-\tanh\frac{{b}}{2})^{N-1} a_0^* a_1
f_N(\tau) &\displaystyle \frac{e^{-\frac{{b}}{2}}}{2 \cosh\frac{{b}}{2}} - \frac{1}{2}\left(
\frac{e^{-\frac{{b}}{2}}}{\cosh\frac{{b}}{2}}- 2|a_1|^2\right) |f_N(\tau)|^2 
\end{array}
\right),
\end{eqnarray}
where
\begin{eqnarray}\label{f}
f_N(\tau)= \sum_{k=1}^N e^{ i \varepsilon_k \tau} g_{1k} g_{Nk}.
\end{eqnarray}
In our calculations, 
we use $f_N$ as a global characteristic of the transmission line and represent it in the form
\begin{eqnarray}\label{fRPhi}
f_N(\tau)= R_N(\tau) e^{2 i \pi \Phi_N(\tau)},
\end{eqnarray}
where $R_N$ and $\Phi_N$  are the amplitude and the phase of $f_N$, respectively. Then, eq.(\ref{ArhoN}) 
reduces into eq.(\ref{rhoN0}). The maximal creatable region corresponds to the maximum  of $R_N(\tau)$.
This maximum $R(N)$  together with the 
appropriate time instant $\tau_{max}(N)$  are found as functions of the chain 
length $N$ in Sec.\ref{Section:RN}, 
see Fig.\ref{Fig:FT}.

\section{Appendix B: The asymptotic behavior of function $R_N(\tau)$ as $N\to \infty$.}

Let us  rewrite the function $f_N(\tau)$ for odd $N$ as (the case of even $N$ can be treated similarly) 
\begin{equation}\label{eq1}
f_N(\tau)=
\frac{2}{N+1}\sum_k e^{i\epsilon _k \tau}\sin (kN) \sin(k)= 
\frac{2}{N+1}\sum_{n=1}^N \sin^2 \left(\frac{\pi n}{N+1}  \right)\cos \left[\tau\cos\left(\frac{\pi n}{N+1}\right)\right]
\end{equation}
%Below we use the dimensionless time $\tau=Dt$. 
We can introduce the Bessel functions into eq.(\ref{eq1}) using the following well known relation
\cite{Abramowitz}:
\begin{equation}\label{eq2}
\cos \left[\tau\cos\left(\frac{\pi n}{N+1}\right)\right]=J_0(\tau)+2\sum_{m=1}^\infty (-1)^m J_{2m}(\tau) \cos\left(\frac{2\pi mn}{N+1}\right).
\end{equation}
Substituting eq.\eqref{eq2} into eq.\eqref{eq1} we  obtain:
\begin{eqnarray}\label{eq3}
&&
f_N(\tau)=
\\\nonumber
&&
-\sum_{p=0}^\infty (-1)^{(2p+1)\frac{N+1}{2}}
\left(
2  J_{(2p+1)(N+1)}(\tau) 
+ J_{(2p+1)(N+1)-2}(\tau) +
J_{(2p+1)(N+1)+2}(\tau)\right).
\end{eqnarray}
It can be simply verified using the numerical simulation that the  behavior of $f_N(\tau)$ over the time interval $0<\tau\lesssim 2 N$
with $N>2$ is  governed  by the first term in the above sum over $p$. In other words, we have for 
the amplitude of $f_N$:
\begin{eqnarray}\label{eq3appr}
R^{appr}_N(\tau)&=& \Big | J_{N+3}(\tau) 
+J_{N-1}(\tau) 
+2 J_{N+1}(\tau)\Big|.
\end{eqnarray}
Being derived for odd $N$, this formula holds for even $N$ as well, giving  the  maximum $R$ in formulas 
(\ref{IJ}) and (\ref{Jdef}) with the accuracy increasing with $N$. Thus,  $|R(3)-R^{appr}(3)| \sim 0.001$ 
and  $|R(5)-R^{appr}(5)| \sim 10^{-5}$. Consequently, although  formula (\ref{eq3appr}), generally speaking, 
assumes large $N$, it is applicable to the short chains as well.


\begin{thebibliography}{99}
 
 
\bibitem{Bose}
S. Bose: Quantum Communication through an Unmodulated Spin Chain, Phys. Rev. Lett., {\bf   91}, 207901 (2003)

\bibitem{CDEL}
 M.Christandl, N.Datta, A.Ekert and A.J.Landahl: Perfect State Transfer in Quantum Spin Networks, 
 Phys.Rev.Lett. {\bf   92}, 187902 (2004)

\bibitem{ACDE}
 C.Albanese, M.Christandl, N.Datta and A.Ekert:
 Mirror Inversion of Quantum States in Linear Registers, Phys.Rev.Lett. {\bf   93}, 230502 (2004)

\bibitem{KS}
 P.Karbach and J.Stolze: Spin chains as perfect quantum state mirrors, Phys.Rev.A {\bf   72}, 030301(R) (2005)

 
\bibitem{GKMT}
 G.Gualdi, V.Kostak, I.Marzoli and P.Tombesi: Perfect state transfer in long-range interacting spin chains, 
 Phys.Rev. A {\bf   78}, 022325 (2008)

 
 
\bibitem{WLKGGB}
A.W\'ojcik, T.Luczak, P.Kurzy\'nski, A.Grudka, T.Gdala, and M.Bednarska: Unmodulated spin chains as universal quantum wires,
Phys. Rev. A {\bf   72}, 034303 (2005)
 
 
 
\bibitem{CRMF}
G. De Chiara, D. Rossini, S. Montangero, R. Fazio: From perfect to fractal transmission in spin chains, 
Phys. Rev. A {\bf   72}, 012323 (2005)

\bibitem{ZASO}
 A. Zwick, G.A. \'Alvarez,
J. Stolze, O. Osenda: 
Robustness of spin-coupling distributions for perfect quantum state transfer, Phys. Rev. A {\bf   84}, 022311 (2011)

 
 \bibitem{ZASO2}
 A. Zwick, G.A. \'Alvarez,
J. Stolze, O. Osenda: Spin chains for robust state transfer: Modified boundary couplings versus completely engineered chains, 
Phys. Rev. A {\bf   85}, 012318 (2012)

\bibitem{ZASO3}
A. Zwick, G.A. \'Alvarez, J. Stolze, O Osenda: 
Quantum state transfer in disordered spin chains: How much engineering is reasonable?,
Quant. Inf. Comput. {\bf 15}, 582-600 (2015)

 

 \bibitem{SAOZ}
 J.Stolze, G. A. \'Alvarez,
O. Osenda, A. Zwick in
 Quantum State Transfer and Network Engineering.
Quantum Science and Technology,
ed. by  G.M.Nikolopoulos and I.Jex, Springer Berlin Heidelberg, Berlin, 149-182  (2014) 


\bibitem{KZ_2008}
E.I. Kuznetsova, A.I. Zenchuk: High-probability quantum state transfer in an alternating open spin chain
with an X Y Hamiltonian, Phys. Lett. A {\bf 372}, 6134-6140  (2008) 



 

\bibitem{NJ}
G.M.Nikolopoulos and I.Jex, eds.: Quantum State Transfer and Network Engineering, Series in
Quantum Science and Technology, Springer, Berlin Heidelberg (2014)


 


\bibitem{BK}
A. Bayat and V. Karimipour: Thermal effects on quantum communication through spin chains, 
Phys.Rev.A {\bf 71}, 042330 (2005)


\bibitem{C}
P. Cappellaro: Coherent-state transfer via highly mixed quantum spin chains,
Phys.Rev.A {\bf 83}, 032304 (2011)


\bibitem{QWL}
W. Qin,  Ch. Wang,   G. L. Long: High-dimensional quantum state transfer through a quantum spin chain, 
Phys.Rev.A {\bf 87}, 012339 (2013)

 
 
\bibitem{B}
A.Bayat: Arbitrary perfect state transfer in d-level spin chains,
Phys. Rev. A {\bf 89}, 062302 (2014)


 
\bibitem{JKSS}
C. Godsil, S. Kirkland, S. Severini,  Ja. Smith: Number-Theoretic Nature of Communication in Quantum Spin Systems
Phys. Rev. Lett. {\bf 109}, 050502 (2012)


\bibitem{SO}
R.Sousa, Ya. Omar: Pretty good state transfer of entangled states through quantum spin chains,
New J. Phys. {\bf 16}, 123003 (2014).

 
 
 \bibitem{BB}
 D. Burgarth and S. Bose:Conclusive and arbitrarily perfect quantum-state transfer using parallel spin-chain channels,
Phys.Rev.A {\bf 71}, 052315 (2005)

\bibitem{SJBB}
K. Shizume,  K. Jacobs,  D. Burgarth,  and S. Bose: Quantum communication via a continuously monitored dual spin chain,
Phys. Rev. A {\bf 75}, 062328 (2007)
 
 
\bibitem{SKKVP}
M. Sandberg, E. Knill, E. Kapit, M.R.Vissers, D.P.Pappas: 
Efficient quantum state transfer in an engineered chain of quantum bits, to appear in 
Quant. Inf. Proc, 2015:1-12./DOI 10.1007/s11128-015-1152-4


\bibitem{LS}
P. Lorenz,  J. Stolze: Transferring entangled states through spin chains by boundary-state multiplets,
Phys. Rev. A {\bf 90}, 044301 (2014)

 
\bibitem{BBVB}
L.Banchi,  A. Bayat,  P. Verrucchi, and S.Bose: 
Nonperturbative Entangling Gates between Distant Qubits Using Uniform Cold Atom Chains, 
Phys.Rev.Let. {\bf 106}, 140501 (2011)
 
 
 \bibitem{QWZ}
 W. Qin, Ch. Wang, and X. Zhang: 
 Protected quantum-state transfer in decoherence-free subspaces, Phys. Rev. A {\bf 91}, 042303
(2015).
 
 
\bibitem{ZZHE}
M.Zukowski, A.Zeilinger, M.A.Horne, A.K.Ekert: 
''Event-ready-detectors'' Bell experiment via entanglement swapping, Phys. Rev.
Lett. {\bf 71}, 4287-4290 (1993)

 
\bibitem{BPMEWZ}
D.Bouwmeester, J.-W. Pan, K.Mattle, M.Eibl, H.Weinfurter, and  A. Zeilinger:
Experimental quantum teleportation, 
Nature {\bf 390}, 575-579  (1997)

\bibitem{BBMHP}
D. Boschi,  S. Branca,  F. De Martini, L. Hardy,  and S. Popescu:
Experimental Realization of Teleporting an Unknown Pure 
Quantum State via Dual Classical and Einstein-Podolsky-Rosen Channels,
Phys. Rev. Lett. {\bf 80}, 1121-1125 (1998)

 
\bibitem{PBGWK2}
N.A.Peters, J.T.Barreiro,  M.E.Goggin, T.-C.Wei,  and P.G.Kwiat: 
Remote State Preparation: Arbitrary Remote Control of Photon Polarization, Phys.Rev.Lett. {\bf   94}, 
150502 (2005) 

\bibitem{PBGWK}
N.A.Peters, J.T.Barreiro, M.E.Goggin, T.-C.Wei, and P.G.Kwiat: 
Remote state preparation: arbitrary remote control of photon polarizations for quantum communication,
in  Quantum
Communications and Quantum Imaging III, ed. R.E.Meyers,
Ya.Shih, Proc. of SPIE {\bf   5893} 589301 (SPIE, Bellingham, WA, 2005) 


\bibitem{DLMRKBPVZBW}
B.Dakic, Ya.O.Lipp, X.Ma, M.Ringbauer, S.Kropatschek,
S.Barz, T.Paterek, V.Vedral, A.Zeilinger, C.Brukner, and P.Walther: 
Quantum discord as resource for remote state preparation, 
Nat. Phys. {\bf   8}, 666-670 (2012). 

\bibitem{XLYG}
G.Y. Xiang, J.Li, B.Yu, and G.C.Guo: Remote preparation of mixed states via noisy entanglement,
Phys. Rev. A {\bf   72}, 012315  (2005)



\bibitem{PSB}
S.Pouyandeh,  F. Shahbazi,  A. Bayat: Measurement-induced dynamics for spin-chain quantum
communication and its application for optical lattices,
Phys.Rev.A {\bf 90}, 012337 (2014)


\bibitem{YBB}
S.Yang,  A. Bayat, S. Bose: Entanglement-enhanced information transfer through strongly
correlated systems and its application to optical lattices, Phys.Rev.A {\bf 84}, 020302 (2011)



 
\bibitem{Z_2012}
A.I.Zenchuk: Information propagation in a quantum system: examples of open spin-1/2 chains,
J. Phys. A: Math. Theor. {\bf   45} 115306 (2012) 
 
\bibitem{PS}
S. Pouyandeh, F. Shahbazi: Quantum state transfer in XXZ spin chains: A measurement induced transport method, 
Int. J. Quantum Inform. {\bf 13}, 1550030 (2015)


\bibitem{Z_2014}
A.I.Zenchuk: Remote  creation of a one-qubit mixed state through a short homogeneous spin-1/2 chain,
Phys. Rev. A {\bf 90}, 052302 (2014) 


\bibitem{BZ_2015}
G.A.Bochkin, A.I.Zenchuk: Remote one-qubit-state control using the pure initial state of a two-qubit sender:
Selective-region and eigenvalue creation,  Phys.Rev.A {\bf 91}, 062326 (2015)
 
 
 

\bibitem{FBE}
E.B.Fel'dman, R.Br\"uschweiler and R.R.Ernst:
From regular to erratic quantum dynamics in long spin 1/2 chains with an XY Hamiltonian, 
Chem.Phys.Lett. {\bf 294}, 297-304
(1998)

\bibitem{Abragam}
A. Abragam, ''The Principles of Nuclear Magnetism'', Oxford, Clarendon Press (1961)

\bibitem{BMGP}
J. Baum, M. Munowitz, A. N. Garroway, and A. Pines:
Multiple-quantum dynamics in solid state NMR, J. Chem. Phys.
{\bf 83}, 2015-2025 (1985)

\bibitem{OBSFA}
 I.S. Oliveira, T. J. Bonagamba, R.S. Sarthour, J. C.C. Freitas and E. R. deAzevedo:
NMR Quantum Information Processing, Elsevier, Amsterdam, 2007

\bibitem{F}
E.B.Fel'dman: Multiple quanum NMR in one dimensional and nano-scale systems: theory and computer simulations, 
Appl.Magn.Res. {\bf 45}, 797-806 (2014)



 \bibitem{Goldman}
M.Goldman: Spin temperature and nuclear magnetic resonance in solids,
Clarendon Press. Oxford, 1970



\bibitem{JW}
P.Jordan, E.Wigner: \"Uber das Paulische \"Aquivalenzverbot, Z.Phys. {\bf 47}, 631-651 (1928)

 
\bibitem{CG}
H.B.Cruz, L.L.Goncalves: 
Time-dependent correlations of the one-dimensional isotropic XY model, J. Phys. C: Solid State Phys. {\bf 14}, 2785-2791 (1981)

 
\bibitem{Abramowitz}
M.A. Abramowitz, I.A. Stegun, Handbook of Mathematical Functions with Formulas,
Graphs, and Mathematical Tables. National bureau of standards. Applied Mathematical series-55, 355-389 (1964).

\bibitem{GRTZ}
N.Gisin, G. Ribordy, W.Tittel,   H.Zbinden: Quantum cryptography, 
Rev. Mod. Phys. {\bf 74}, 145-195
(2002)

\bibitem{HBB}
M.Hillery, V.Buzek,  A.Berthiaume: Quantum secret sharing ,
Phys.Rev.A {\bf 59}, 1829-1834 (1999)


 \bibitem{SCCS}
S. Sazim, V. Chiranjeevi, I. Chakrabarty, K. Srinathan: Retrieving and routing quantum information in a quantum network
 , Quant.Inf.Proc., 
 {\bf 14},  4651-4664
 (2015)
 
\end{thebibliography}
\end{document}